\documentclass[fleqn,usenatbib]{mnras}

\usepackage{newtxtext,newtxmath}

\usepackage[T1]{fontenc}

\DeclareRobustCommand{\VAN}[3]{#2}
\let\VANthebibliography\thebibliography
\def\thebibliography{\DeclareRobustCommand{\VAN}[3]{##3}\VANthebibliography}

\usepackage{graphicx}	
\usepackage{amsmath}	
\usepackage{widetext}

\usepackage{color}
\usepackage{soul}
\definecolor{vero}{rgb}{0.75, 0.40, 0.73}

\definecolor{jim}{rgb}{1.00, 0.00, 0.00}

\definecolor{tali}{rgb}{0.00, 0.75, 0.5}

\title[Early-type star magnetic forces and torques]{Using ZDI maps to determine magnetic forces and torques at the photospheres of Early-type stars}

\author[James MacDonald et al.]{
James MacDonald$^{1}$\thanks{E-mail: jimmacd@udel.edu (JM)},
Tali Natan$^{1}$, Véronique Petit$^{1}$,
Oleg Kochukhov$^{2}$, Matthew E. Shultz$^{1}$
\\
$^{1}$Dept. Physics and Astronomy, University of Delaware, Newark, DE 19716, USA\\
$^{2}$Dept. Physics and Astronomy, Uppsala University, Box 516, SE-751 20 Uppsala, Sweden
}

\date{Accepted XXX. Received YYY; in original form ZZZ}

\pubyear{2021}

\begin{document}
\label{firstpage}
\pagerange{\pageref{firstpage}--\pageref{lastpage}}
\maketitle

\begin{abstract}
We use the magnetic field components measured by Zeeman Doppler imaging (ZDI) to calculate the stellar surface force and torque due to magnetic stresses for the fast rotators $\sigma$ Ori E, 36 Lyn and CU Vir, and the slow rotator $\tau$ Sco. If we assume the stars have spherical photospheres, the estimated torques give spin down time scales no larger than $7 \times 10^5$ yr. For $\sigma$ Ori E, the predicted spin down time scale, $\simeq 6000$ yr, is much less than the observationally measured time scale of $\simeq 10^6$ yr. However, for CU Vir, we find that the spin down time scale from its ZDI map is $7 \times 10^5$ yr in good agreement with its average rate of spin down from 1960 to 2010.  

With the exception of $\tau$ Sco, the net force due to magnetic stresses at the stellar surface are large compared to the surface-integrated pressure. We discuss possible reasons for the large values of the forces (and torques), and suggest that the likely explanation is that rotation and the magnetic stresses create significant departures from spherical symmetry. 
\end{abstract}




\section{Introduction}

Spectroscopic measurements show that most massive main sequence stars are fast rotators whereas low mass main sequence stars of spectral type later than F5 rotate slowly \citep{1955ApJ...121..653S}. Rotation period measurements with the Kepler satellite show that the transition from fast to slow rotation occurs in the spectral type range F2 - F5 \citep{2013A&A...557L..10N}. \citet{1962AnAp...25...18S} proposed the widely accepted explanation that this dichotomy is due to the low mass stars having dynamo-generated magnetic fields that force their stellar winds to co-rotate out to a critical Alfvén radius, which results in angular momentum loss rates that lead to slow rotation. In contrast, even though they have powerful winds, most massive stars do not have magnetic fields strong enough for significant angular momentum loss during their short lifetimes. However, a few percent of massive stars do have strong surface magnetic fields \citep{2009ARA&A..47..333D, 2017MNRAS.465.2432G}, detected by Zeeman splitting and spectropolarimetric observations. 

Consistent with the expectation that a magnetically trapped wind should rapidly reduce angular momentum, magnetic hot stars have systematically lower projected rotational velocities than non-magnetic stars with similar spectral types \citep{2018MNRAS.475.5144S, 1995ApJS...99..135A, 2009ARA&A..47..333D}. Furthermore, the rotational periods of magnetic stars increase over time \citep{2019MNRAS.490..274S}, with young stars having rotational periods as short as 0.5 d  \citep[e.g.][]{2012MNRAS.419.1610G,2013MNRAS.429..177R}, whereas older stars can have rotational periods of decades \citep[e.g.][] {2017MNRAS.471.2286S, 2021MNRAS.506.2296E}. 
In a small number of cases, rotational period change has been directly measured. The results of such studies have been mixed. The canonical magnetic star $\sigma$ Ori E slows down at approximately the rate predicted due to wind magnetic braking \citep[][Petit et al., in prep.]{2010ApJ...714L.318T}. However, other stars have been observed to exhibit spin-up \citep{2019MNRAS.486.5558S}, or even complex cycles of spin-up and spin-down \citep[e.g.][]{2008A&A...485..585M, 2011A&A...534L...5M,2017ASPC..510..220M}. 

Because the torque predicted from the standard magnetic braking of a stellar wind model gives rise to a steady loss of angular momentum that is inconsistent with rotation period decreases over short timescales, in this paper we investigate constraints on the torque due to magnetic stresses acting at the stellar surface obtained by direct integration of ZDI magnetic field maps. In the next section, we briefly review the theory of angular loss due to magnetic braking of stellar winds. In section 3, we derive integral expressions for the force and torque on the star due to the magnetic stresses at the stellar surface, and in section 4 we evaluate these integral expressions for the components of the force and torque by using the field components from ZDI maps of $\sigma$ Ori E, CU Vir, $\tau$ Sco, and 36 Lyn, assuming that the stellar surface is spherical. Discussion and conclusions are given in sections 5 and 6, respectively.

\section{The spin down torque}

The theory of angular momentum loss due to magnetically braked stellar winds has been developed in a number of papers \citep[e.g.][]{1967ApJ...148..217W, 1968MNRAS.138..359M, 1984LNP...193...49M, 1976ApJ...210..498B, 1985A&A...152..121S, 1987MNRAS.226...57M, 1988ApJ...333..236K, 2008ApJ...678.1109M, 2009MNRAS.392.1022U, 2012ApJ...754L..26M, 2012ApJ...746...43R, 2015ApJ...798..116R}. In addition to the stellar angular velocity and mass loss rate, the angular momentum loss rate depends on the strength and topology of the magnetic field. Regions in which the flow speed, $v$, is less than the Alfvén speed, $v_{A}$, are forced to co-rotate with the star. In the simple case of a monopole field, for a slow rotator the angular momentum loss rate is
\begin{equation} \label{eq:1}
    \dot{J}=\frac{2}{3}\dot{M}{{R}_{A}}^{2}\Omega,
\end{equation}
where $\dot{M}$ is the mass loss rate, $\Omega$ is the stellar angular velocity and $R_{A}$ is the radius of the spherical Alfvén surface that bounds the co-rotating wind region. From mass conservation for a steady wind and flux conservation for a magnetic monopole, it is straightforward to show that
\begin{equation} \label{eq:2}
    {{R}_{A}}^{2}=\frac{{{R}_{*}}^{4}{{B}_{*}}^{2}}{\left| {\dot{M}} \right|{{v}_{A}}},
\end{equation}
where the asterisk denotes quantities at the stellar surface and $v_{A}$ is the wind speed at the Alfvén surface. 

A number of works use equation (\ref{eq:1}) for the angular momentum loss rate by modifying the expression for $R_{A}$ to account for differences in the field geometry and wind driving mechanism (see e.g. \cite{2009MNRAS.392.1022U}, \cite{2015ApJ...798..116R}). Essentially ${R_{A}}^2$ is replaced by $\langle {R_{A}}^2 \rangle$, its mass-loss weighted average over the Alfvén surface as described by \cite{1993MNRAS.262..936W}. 

From comparison with the results of their MHD calculations \citep{2008MNRAS.385...97U}, \cite{2009MNRAS.392.1022U} find for radiatively driven winds in an aligned rotator, the angular momentum loss rate due to a magnetically braked wind is
\begin{equation} \label{eq:3}
 \dot{J}_{mbw}=\frac{2}{3}\dot{M}\Omega {{\left\langle {{{R}_{A}}^{2}} \right\rangle }}=\frac{2}{3}\dot{M}\Omega {{R}_{*}}^{2}{{\left[ 0.29+{{\left( {{\eta }_{*}}+0.25 \right)}^{{1}/{4}\;}} \right]}^{2}},
\end{equation}
where the wind magnetic confinement parameter is defined by \citep{2002ApJ...576..413U}
\begin{equation}  \label{eq:4}
 {{\eta }_{*}}=\frac{{{B}_{eq}}^{2}{{R}_{*}}^{2}}{\left| {\dot{M}} \right|{{v}_{\infty }}}.
\end{equation}
Here $B_{eq}$ is the surface dipole equatorial field strength and ${{v}_{\infty }}$ is the terminal flow speed. 

From their MHD models for a thermally driven wind, \cite{2015ApJ...798..116R} find 
\begin{equation} \label{eq:5}
 \frac{\left\langle {{R}_{A}} \right\rangle }{{{R}_{*}}}={{K}_{1}}{{\left[ \frac{\Upsilon }{{{\left( 1+{{{f}^{2}}}/{{{K}_{2}}^{2}}\; \right)}^{{1}/{2}\;}}} \right]}^{m}},
\end{equation}
where
\begin{equation} \label{eq:6}
 \Upsilon =\frac{{{B}_{eq}}^{2}{{R}_{*}}^{2}}{\left| {\dot{M}} \right|{{v}_{esc}}}
\end{equation}
is the magnetization parameter \citep{2008ApJ...678.1109M} and $f$ is the ratio between the star’s equatorial surface rotation rate and Keplerian speed
\begin{equation} \label{eq:7}
 f=\frac{{{\Omega }}{{R}_{*}}^{{3}/{2}\;}}{{{\left( G{{M}_{*}} \right)}^{{1}/{2}\;}}}.
\end{equation}
For a dipole field, \cite{2015ApJ...798..116R} find $K_{1} = 2.0 \pm 0.1$, $K_{2} = 0.2 \pm 0.1$, $m = 0.235 \pm 0.007$.

For the rest of the paper, we concentrate on radiatively driven winds. Comparison of the theoretical wind torque, given by equation (\ref{eq:3}) and the relevant expression for $R_{A}$, with observed spin down rates requires that the mass loss rate and other quantities are known. Also the derivations of these expressions are based on the assumption that the magnetic and rotation axes are aligned, which may not be the case. Indeed, \cite{2019MNRAS.490..274S} find from their analysis of a sample of magnetic early B-type stars that the obliquity angle is statistically consistent with a random distribution. \cite{2013MNRAS.429..398P} provide a detailed discussion of the  correlated errors in determining the angular momentum loss rate from equation (\ref{eq:3}). As yet there has been no modification to (\ref{eq:3}) to take into account the effect of non-zero obliquity but \cite{2022MNRAS.515..237S} have performed 3D MHD calculations for a few cases of an obliquie rotator, which indicate that the variation in angular momentum loss rate is of order 10 - 20 \%.

Because of these limitations, we explore an alternative approach in which the magnetic torque is evaluated from the observed magnetic field components. In the next section, we derive expressions for the net force and torque in terms of integrals of the magnetic stresses over the stellar surface.

\section{Force and torque from magnetic stresses}
The force on the star from magnetic stresses is given by
\begin{equation} \label{eq:8}
    \mathbf{F}=-\frac{1}{4\pi }\int\limits_{S}{\frac{{{B}^{2}}}{2}\mathbf{n}-\left( \mathbf{n}\cdot \mathbf{B} \right)\mathbf{B}dS.}
\end{equation}

The torque on the star resulting from just the magnetic stresses is then
\begin{equation} \label{eq:9}
 \mathbf{\tau }=-\frac{1}{4\pi }\int\limits_{S}{\frac{{{B}^{2}}}{2}\left( \mathbf{r}\times \mathbf{n} \right)-\left( \mathbf{r}\times \mathbf{B} \right)\left( \mathbf{n}\cdot \mathbf{B} \right)dS.}
\end{equation}

In the particular case of a spherical star, the force and torque expressions simplify to
\begin{equation} \label{eq:10}
 \mathbf{F}=-\frac{{{R}_{*}}^{2}}{4\pi }\int\limits_{S}{\left( \frac{{{B}^{2}}}{2}\mathbf{\hat{r}}-{{B}_{r}}\mathbf{B} \right)\sin \theta d\theta d\phi .}
\end{equation}
and
\begin{equation} \label{eq:11}
 \mathbf{\tau}=\frac{{{R}_{*}}^{3}}{4\pi }\int\limits_{S}{\left( \mathbf{\hat{r}}\times \mathbf{B} \right){{B}_{r}}\sin \theta d\theta d\phi.}
\end{equation}

Taking the rotation axis as the $z$-axis, the components of the force and torque on a spherical star are
\pagebreak
\begin{widetext}
\begin{equation} \label{eq:12}
  \begin{split}
  & {{F}_{x}}=-\frac{{{R}_{*}}^{2}}{8\pi }\int\limits_{S}{\left[ \left( {{B}_{\theta }}^{2}+{{B}_{\phi }}^{2}-{{B}_{r}}^{2} \right)\sin \theta \cos \phi-2{{B}_{r}}{{B}_{\theta }}\cos \theta \cos \phi +2{{B}_{r}}{{B}_{\phi }}\sin \phi  \right]\sin \theta d\theta d\phi }, \\ 
 & {{F}_{y}}=-\frac{{{R}_{*}}^{2}}{8\pi }\int\limits_{S}{\left[ \left( {{B}_{\theta }}^{2}+{{B}_{\phi }}^{2}-{{B}_{r}}^{2} \right)\sin \theta \sin \phi -2{{B}_{r}}{{B}_{\theta }}\cos \theta \sin \phi -2{{B}_{r}}{{B}_{\phi }}\cos \phi  \right]\sin \theta d\theta d\phi }, \\ 
 & {{F}_{z}}=-\frac{{{R}_{*}}^{2}}{8\pi }\int\limits_{S}{\left[ \left( {{B}_{\theta }}^{2}+{{B}_{\phi }}^{2}-{{B}_{r}}^{2} \right)\cos \theta +2{{B}_{r}}{{B}_{\theta }}\sin \theta  \right]\sin \theta d\theta d\phi }. \\ 
  \end{split}
\end{equation}
\end{widetext}
and
\begin{equation} \label{eq:13}
	\begin{split}
  & {{\tau }_{x}}=-\frac{{{R}_{*}}^{3}}{4\pi }\int\limits_{S}{\left( {{B}_{\theta }}\sin \phi +{{B}_{\phi }}\sin \theta \cos \phi  \right){{B}_{r}}\sin \theta d\theta d\phi }, \\ 
 & {{\tau }_{y}}=\frac{{{R}_{*}}^{3}}{4\pi }\int\limits_{S}{\left( {{B}_{\theta }}\cos \phi -{{B}_{\phi }}\sin \theta \sin \phi  \right){{B}_{r}}\sin \theta d\theta d\phi }. \\ 
 & {{\tau }_{z}}=-\frac{{{R}_{*}}^{3}}{4\pi }\int\limits_{S}{{{\sin }^{2}}\theta {{B}_{r}}{{B}_{\phi }}d\theta d\phi }. \\
   \end{split}
\end{equation}

A more general expression for the total torque on a non-spherical rotating magnetic star that includes the contribution from the stellar wind and thermal pressure has been derived by \cite{1970MNRAS.149..197M}. \cite{2014MNRAS.438.1162V} used this result to determine the angular momentum evolution of early-M dwarf stars using 3D numerical simulations of the stellar wind which apply the surface field geometry from ZDI maps as an interior boundary condition.

In the limit that the stellar radius is much smaller than the Alfv\'{e}n radius, \cite{1967ApJ...148..217W} show that in their equatorial plane analysis in which the poloidal field is described by a pure monopole, the radial and toroidal components of the field at the stellar surface are related by 
\begin{equation} \label{eq:14}
	{{B}_{\phi }}=-\frac{\Omega {{R}_{*}}}{{{u}_{a}}}{{B}_{r}}.
\end{equation} 
where $u_a$ is the Alfv\'{e}n speed at the Alfv\'{e}n radius. \cite{1969SSRv....9..325P} has used a generalization of this result to the whole stellar surface to find an expression for the torque
\begin{equation} \label{eq:15}
	\tau_{mono} = -\frac{2}{3}\frac{\Omega {{R}_{*}}}{{{u}_{a}}}{{R}_{*}}^{3}{{B}_{*}}^{2},
\end{equation}	
where $B_{*}$  is the strength of the monopolar radial field at the stellar surface .

In this simple monopole model, the toroidal magnetic field always points in the direction that leads to a torque that reduces the angular momentum of the star. As we will find from analysis of the ZDI maps for stars with more complicated field topologies, the toroidal field can locally be in the opposite direction and give a positive contribution to the magnetic torque surface integral.

\section{Observationally measured forces and torques}
We have used the ZDI maps (shown in the Appendix) of the magnetic fields of $\sigma$ Ori E \citep{2015MNRAS.451.2015O}, CU Vir \citep{2014A&A...565A..83K}, $\tau$ Sco \citep{2016A&A...586A..30K} and 36 Lyn \citep{2018MNRAS.473.3367O} to estimate the forces and torques acting on the stars from magnetic stresses. The integrals in equations (12) and (13) are approximated by sums of form, e.g.

\begin{equation} \label{eq:16}
\int\limits_{S}{\left( {{B}_{r}}{{B}_{\phi }}\sin \theta  \right)\sin \theta d\theta d\phi }\approx \sum\limits_{k}{{{\left( {{B}_{r}}{{B}_{\phi }}\sin \theta  \right)}_{k}}\sin {{\theta }_{k}}d{{\theta }_{k}}d{{\phi }_{k}}},
\end{equation}

where $d{\phi}_k$ and $d{\theta}_k$ are determined from the grid spacing. The number of points in the map is 1176 for $\sigma$ Ori E and $\tau$ Sco or 1876 for 36 Lyn and CU Vir.  We test the accuracy of the integral approximation by using much higher resolution maps for CU Vir created from the spherical harmonic series expansions for the field components given in \cite{2014A&A...565A..83K}. We find that the errors from using the low resolution map are small with the maximum difference in any of the force and torque components of $4 \times 10^{25}$ dyn and $2 \times 10^{36}$ dyn cm, respectively. The relative errors in the magnitudes of the force and torque are 0.4\% and 0.07\%, respectively.

We have also down scaled to 1176 grid points to estimate numerical errors for $\sigma$ Ori E and $\tau$ Sco. The maximum difference in any of the force and torque components is now $9 \times 10^{25}$ dyn and $3 \times 10^{36}$ dyn cm, respectively. The relative errors in the magnitudes of the force and torque are 0.6\% and 0.1\%, respectively.

Our results from integration of the low resolution maps are summarized in Table \ref{tab:force and torque}. The final column gives the predicted spin down time scale, $t_{ZDI}$, assuming that the star rotates like a solid body.

\begin{table*}
\centering
\caption{Components of the force and torque calculated from ZDI maps of $\sigma$ Ori E, CU Vir, $\tau$ Sco and 36 Lyn, assuming spherical stars }
\label{tab:force and torque}
\setlength{\tabcolsep}{4pt}
\begin{tabular}{l|ccccccc}  
\hline
Star & $F_x$ &  $F_y$  & $F_z$ & $\tau_x$ & $\tau_y$ &  $\tau_z$ & $t_{ZDI}$ \\ 
             & (dyn)     & (dyn)  & (dyn)  & (dyn cm)      & (dyn cm)    & (dyn cm)	 &  ($10^3$ yr)	 \\ \hline
 $\sigma$ Ori E     &  1.20 $\times 10^{29}$ &  7.88 $\times 10^{28}$ &  8.52 $\times 10^{27}$ & -6.08 $\times 10^{40}$ &  1.91 $\times 10^{40}$ & -3.29 $\times 10^{40}$ & 5.9 \\
 CU Vir             & -4.80 $\times 10^{27}$ &  4.53 $\times 10^{26}$ &  1.02 $\times 10^{28}$ &  6.29 $\times 10^{38}$ &  4.80 $\times 10^{38}$ & -5.9 $\times 10^{37}$ & 710 \\
 $\tau$ Sco         & -1.76 $\times 10^{27}$ & -1.13 $\times 10^{27}$ & -1.29 $\times 10^{26}$ &  2.56 $\times 10^{38}$ & -5.95 $\times 10^{38}$ & -3.87 $\times 10^{38}$ & 66 \\
 36 Lyn             & -4.70 $\times 10^{28}$ &  1.95 $\times 10^{28}$ &  2.16 $\times 10^{28}$ & -6.16 $\times 10^{39}$ & -1.32 $\times 10^{40}$ & -1.47 $\times 10^{39}$ & 12 \\
\hline
\end{tabular}
\centering						       
\end{table*}

For comparison purposes, we give in Table 2 the torque ($\tau_{mbw}$) and rigid body spin down time scale ($t_{mbw}$) estimated by using equation (\ref{eq:3}). Our adopted stellar parameters are given in columns (2) - (6).

\begin{table*}
\centering
\caption{Torque and spin down time scale from equation (\ref{eq:3}) for $\sigma$ Ori E, CU Vir, $\tau$ Sco and 36 Lyn}
\label{tab:torque and spindown time}
\setlength{\tabcolsep}{4pt}
\begin{tabular}{l|cccccccc}  
\hline
Star & Mass &  Radius & Rotational Inertia  & $B_{eq}$ & $\dot{M}$ & ${v}_{\infty}$ &  $\tau_{mbw}$ & $t_{mbw}$ \\ 
             &  $M_{\odot}$        & $R_{\odot}$ & $10^{55}$ g cm$^2$  &  kG    & $M_{\odot}$ yr$^{-1}$ & km s$^{-1}$       & (dyn cm) &  ($10^3$ yr)	 \\ \hline
 $\sigma$ Ori E     &  8.1 &  3.70 & 11.5 &  4.1 & $1.2 \times 10^{-10}$ &  1200 & -2.4 $\times 10^{37}$ & $8.8 \times 10^3$ \\
 CU Vir             & 3.1 &  2.06 & 0.96 &  1.3 &  $1.2 \times 10^{-12}$ &  630 & -2.66 $\times 10^{35}$ & $1.6 \times 10^5$ \\
 $\tau$ Sco         & 17.0 & 5.44 & 48 & 0.16 &  $9.9 \times 10^{-9}$ & 2800 & -5.59 $\times 10^{35}$ & $4.8 \times 10^4$ \\
 36 Lyn             & 4.0 &  3.5 & 3.0 &  1.3 & $1.04 \times 10^{-11}$ & 860 & -6.33 $\times 10^{35}$ & $2.8 \times 10^4$ \\
\hline
\end{tabular}
\centering						       
\end{table*}

In table 3, we compare the spin down time scales, $t_{mbw} = I \omega / \dot{J}_{mbw}$, with $\dot{J}_{mbw}$ estimated from equation (\ref{eq:3}) and, $t_{ZDI}$, from $\tau_z$ given in equation (\ref{eq:13}) with $t_{obs}$, the spin down time scale from observed period changes. We also give $\tau_{obs}$, the torque required to give the spin down time scale from the observed period changes:
\begin{equation} \label{eq:17}
    \tau_{obs} = \frac{dJ}{dt} = \frac{2\pi I \dot{P}}{P^2}
\end{equation}

\begin{table*}
\centering
\caption{Comparison of theoretical and measured spin down time scales for $\sigma$ Ori E, CU Vir, $\tau$ Sco and 36 Lyn}
\label{tab:theoretical and measured spin down time scales}
\setlength{\tabcolsep}{4pt}
\begin{tabular}{l|cccc}  
\hline
Star & $t_{zdi}$  &  $t_{mbw}$  & $t_{obs}$ & $\tau_{obs}$  \\ 
             &  ($10^3$ yr)         & ($10^3$ yr)   &  ($10^3$ yr)    & dyn cm 	 \\ \hline
 $\sigma$ Ori E &  5.9 &  $8.8 \times 10^3$  &  $1.34 \times 10^3$ & $-1.6 \times 10^{38}$ \\
 CU Vir         &  710 &  $1.6 \times 10^5$ &  560 & $-7.1 \times 10^{37}$ \\
 $\tau$ Sco     &   66 &  $4.8 \times 10^4$ & ---  & --- \\
 36 Lyn         &   12 &  $2.8 \times 10^4$ &  --- & --- \\
\hline
\end{tabular}
\centering						       
\end{table*}

\subsection{\texorpdfstring{$\sigma$}{} Ori E}
$\sigma$ Ori E (HD 37479) is magnetic helium-strong variable star that rotates rapidly with period $P = 1.19$ d \citep{1982A&A...116...64G} and has a magnetic obliquity of $47^{\circ} -59^{\circ}$ \citep{2015MNRAS.451.2015O}. From U-band photometric monitoring over the period 2004 to 2009, \cite{2010ApJ...714L.318T} find evidence for an increasing rotation period on a time scale of $P/\dot{P}=$1.34 Myr. \cite{2013ApJ...769...33T} estimate that $\sigma$ Ori E has radius 3.77 $R_{\odot}$ and effective temperature 22500 K. More recently, \cite{2022A&A...657A..60S} have determined the stellar properties to be log $T_{eff} = 4.352 \pm 0.012$ and log $L/L_{\odot} = 3.50 \pm 0.19$. By computing stellar models that include the effects of radially differential rotation, they conclude that $\sigma$ Ori E is a very young star of age less than 1 Myr and initial mass near 9 $M_{\odot}$. Using the DEUCES code \citep{2023MNRAS.525.4700L}, assuming a non-rotating and non-magnetic main sequence star, we find using the Song et al. stellar parameters that $M_{*} = 8.1 \pm 0.5 M_{\odot}$, the age is $4 \times 10^{6}$ to $2 \times 10^{7}$ yr, and the solid body rotational inertia $I = 0.078 - 0.093 MR^2 = 0.8 - 1.5 \times 10^{56}$ g cm$^{2}$ . For solid body rotation, the observed spin down rate would require a total external torque of $\tau \sim  -1.4 \times 10^{38}$ dyn cm.

The Song et al. stellar parameters can also be matched by models of the star in its pre-main sequence phase. We find from our DEUCES models that the mass lies in the range $7.4 - 9.1 M_{\odot}$, and the age is less than 300,000 yr. These non-rotating pre-main sequence models can be ruled out because they predict spin-up due to stellar contraction on a time scale of $\simeq 10^5$ yr. 

None of these age estimates are consistent with the age of 2 - 3 Myr found by \cite{2008AJ....135.1616S} from main sequence fitting, which may indicate that $\sigma$ Ori E is the product of a stellar merger, which has been proposed as the origin mechanism for magnetic massive stars \citep{2019Natur.574..211S}. However, we stress that the details of the stellar models do not significantly influence our conclusion that the torque estimated from the magnetically braked wind angular mass loss formula, $\tau_{mbw}$, is many orders of magnitude less than that from integration of the ZDI map.

To compare with the torque predicted by equation (\ref{eq:3}), we need the wind mass loss rate and terminal velocity, ${v}_{\infty}$. Using the \cite{1999A&A...350..181V, 2000A&A...362..295V, 2001A&A...369..574V} prescription, the mass loss rate for the stellar parameters found above is $1.2 \times 10^{-10} M_{\odot}$ yr$^{-1}$ and ${v}_{\infty}$ = 1200 km s$^{-1}$. However, the effective temperature is near the lower temperature bistability transition temperature, and if the effective temperature were 500 K lower, the mass loss would be an order of magnitude higher and ${v}_{\infty}$ would be 630 km s$^{-1}$.

From the ZDI data, we find that $<B_{eq}^{2}> \simeq$ 16.7 kG$^{2}$. The wind confinement parameter is then $2.4 \times 10^{5}$ - $1.3 \times 10^{6}$ and the torque from equation (\ref{eq:3}) is ~ $0.24 - 1.1\times 10^{38}$ dyn cm. Hence the torque calculated using the larger mass loss rate is consistent with that deduced from the spin down rate, as predicted by \cite{2009MNRAS.392.1022U}. \cite{2013MNRAS.429..398P} adopted the lower mass loss rate and found $t_{mbw}$ = 4.6 Myr, a factor of 3 larger than that observed by \cite{2010ApJ...714L.318T}. In contrast, Parker's result in equation (\ref{eq:15}) gives a spin down torque of magnitude $~3\times 10^{41}$ dyn cm, which is ~2000 times larger than that required by the estimate from the solid body rotation spin down rate.

The torque calculated from the ZDI data by performing numerically the integral in equation (\ref{eq:13}) is $3.3 \times 10^{40}$ dyn cm. Although lower than found using Parker's equation, this torque is still ~200 times larger than estimated from the solid body rotation spin down rate. For a rotation period of 1.19 d, our measured magnetic torque leads to a solid body rotation spin down time scale of $5.9 \times 10^{3}$ yr. 

\begin{figure}
	\includegraphics[width=\columnwidth]{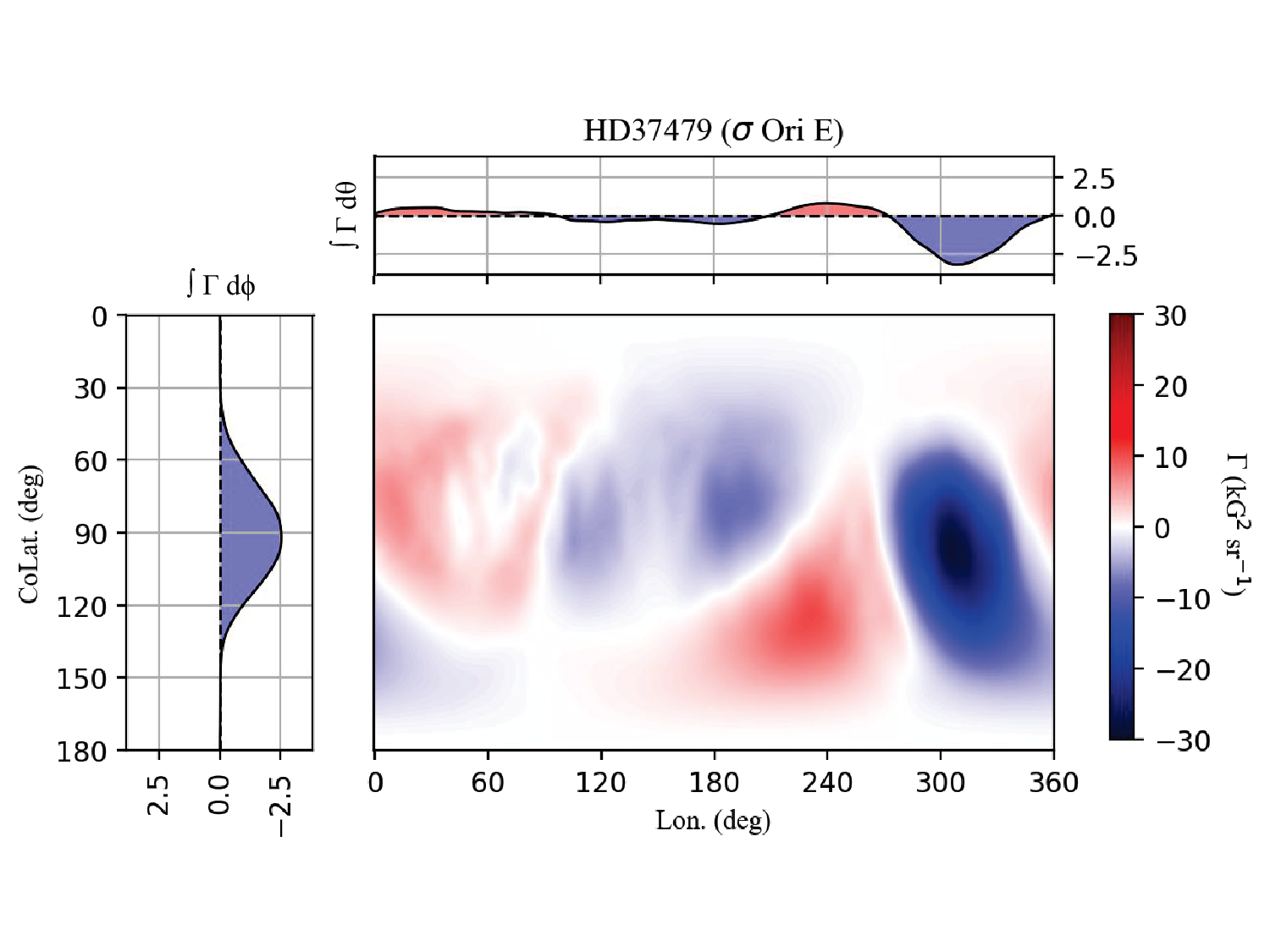}
    \caption{Color plot of $\Gamma$ = $-\sin^{2}\theta B_{r}B_{\phi}$ for $\sigma$ Ori E. $B_{r}$ and $B_{\phi}$ are both in units of kG. Left and top plot show integral of $\Gamma$ with respect to $\phi$ and $\theta$ respectively.}
    \label{fig:figure1}
\end{figure}

In the central panel of fig. 1, we show a surface plot of torque density, $\Gamma = -\sin^{2}\theta B_{r}B_{\phi}$. We see that there is a region of the stellar surface centered near $\phi = 312^{\circ}$, $\theta =100^{\circ}$ where the contribution to the torque integral is large and negative.

Even though Fig. 1 shows that the magnetic torque integrand is not axisymmetric, the left panel shows that when $\Gamma$ is integrated over $\phi$, the contributions to the magnetic torque integral are symmetric about the rotational equator. This symmetry is likely a result of the near-dipole nature of the magnetic field as can be seen from Fig. A1.

\subsection{CU Vir}
CU Vir (HD124224) is a rapidly rotating (P = 0.5207 d) chemically peculiar star that may have discrete and possibly non-monotonic period variations \citep{1998A&A...339..822P, 2013MNRAS.431.2106P, 2011A&A...534L...5M, 2019A&A...625A..34K, 2020PASP..132b4201P}, rather than the smoothly increasing period predicted by magnetic spin down models \citep{2009MNRAS.392.1022U}. From high-resolution spectropolarimetric observations covering an entire rotational period, \cite{2014A&A...565A..83K} find CU Vir’s magnetic field topology deviates significantly from the commonly assumed axisymmetric dipolar configuration. The field is dipolar-like but clearly non-axisymmetric, showing a large difference in the field strength between the regions of opposite polarity.

\cite{2014A&A...565A..83K} adopted $T_{eff}$ = 12750 $\pm$ 250 K based on the SED fitting results of \cite{2004A&A...428..993S} and \cite{2008MNRAS.385..481L} and model atmosphere analysis of spectroscopic data by \cite{1999A&A...348..924K}. Using the $T_{eff}$ value with the observed V magnitude, the Hipparcos trigonometric parallax 12.63 $\pm$ 0.21 mas \citep{2007A&A...474..653V}, and an empirical bolometric correction BC = -0.79 $\pm$ 0.1 \citep{2008MNRAS.385..481L}, Kochukhov et al. found a stellar luminosity $L = 100 \pm 11 L_\odot$ and radius $R = 2.06 \pm 0.14 R_\odot$.

Assuming a main sequence star, we find using our DEUCES models that these stellar parameters are consistent with $M_{*} = 3.12 \pm 0.06 M_{\odot}$, age $1.4 - 11.2 \times 10^{7}$ yr, mass loss rate $1.2 \times 10^{-12} M_{\odot}$ yr$^{-1}$, ${v}_{\infty}$ = 990 km s$^{-1}$ and solid body rotational inertia $I \simeq 0.075 M R^2 \simeq 9.6 \times 10^{54}$ g cm$^{2}$. 

We have estimated the spin torque due to surface magnetic stresses by using the field components from the ZDI measurements \citep{2014A&A...565A..83K}. We find from the ZDI map that $<B_{eq}^{2}>$ = 1.20 kG$^{2}$, and so $\tau_{mbw}$ = $2.7 \times 10^{35}$ dyn cm, which is two orders of magnitude less than found from direct integration of the ZDI map (see table 1). However, the spin-down time scale based on the torque from integration of the ZDI map is in good agreement with the observed spin-down time scale (up to 2010) of $t_{obs} = 6 \times 10^5$ yr \citep{2011A&A...534L...5M}. From rotation periods reported in the literature \citep{1971AJ.....76..630B, 1998A&A...339..822P, 2013MNRAS.431.2106P, 2019A&A...625A..34K}, which we show in figure 2, we estimate a long-term average $\dot{P} = 2.6 \times 10^{-9}$, corresponding to a similar spin-down time scale of $t_{obs} = 5.4 \times 10^5$ yr .

\begin{figure}
	\includegraphics[width=\columnwidth]{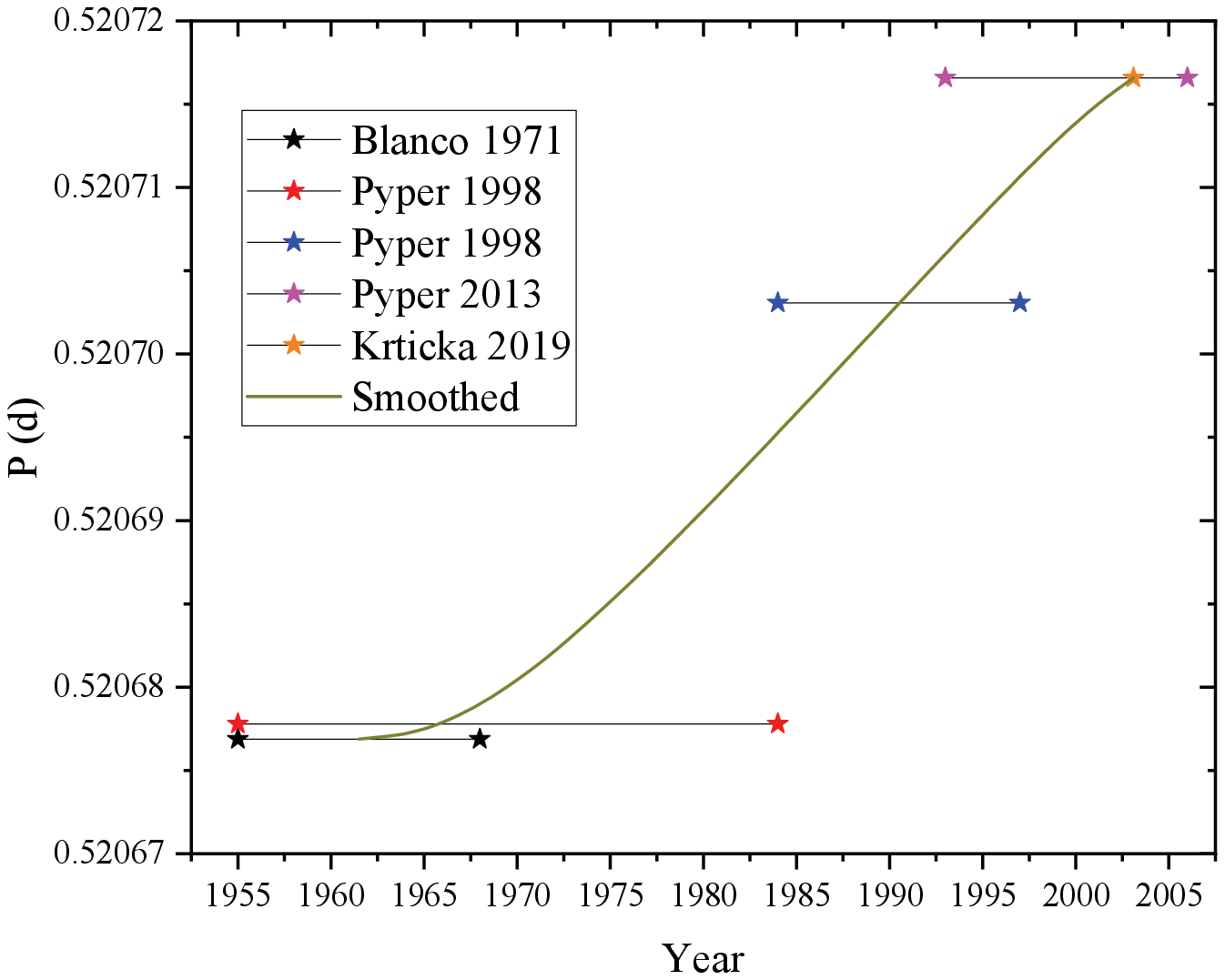}
    \caption{Values for the rotation period of CU Vir taken from the literature. The horizontal lines with symbols at the ends indicate the time span of the observations over which the rotation period was determined. The solid smoothed curve shows the general spin down trend over a span of 50 years.}
    \label{fig:figure2}
\end{figure}

\begin{figure}
	\includegraphics[width=\columnwidth]{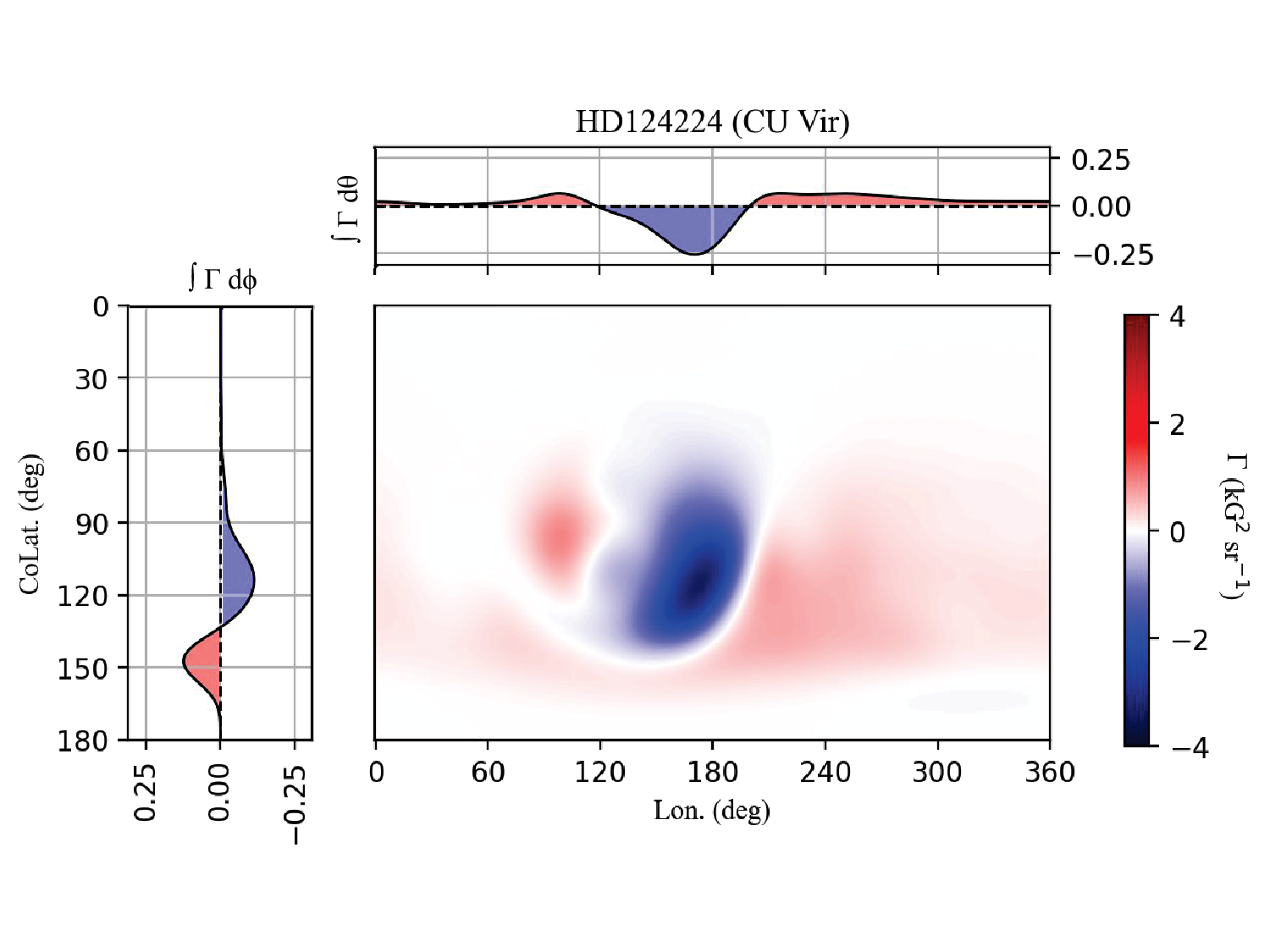}
    \caption{Color plot of $\Gamma$ for CU Vir. $B_{r}$ and $B_{\phi}$ are both in units of kG.}
    \label{fig:figure3}
\end{figure}

Fig. 3 is the same as Fig. 1 except for CU Vir. Again, we see that there is a region of the stellar surface where the contribution to the torque integral is large and negative. However, inspection of the left panel of fig. 2 shows that, unlike in the case of $\sigma$ Ori E, contributions to the magnetic torque integral are not symmetric with respect to the rotational equator. Also, there is a longitudinal band in which the contribution to the net torque is positive, in addition to a band that gives a more dominant negative contribution to the net torque.

\subsection{\texorpdfstring{$\tau$}{} Sco}

$\tau$ Sco (HD149438) of spectral type B0.2V is a relatively slow rotator with P = 41.03 d \citep{2006MNRAS.370..629D}.  It has a complex surface magnetic field topology \citep{2006MNRAS.370..629D, 2009ARA&A..47..333D, 2019MNRAS.490..274S, 2019MNRAS.485.1508S} that deviates significantly from a simple dipole field. \cite{2016A&A...586A..30K} have inverted the circular (Stokes V) spectropolarimetic data to construct possible magnetic field maps. Further linear polarization measurements are needed to produce a unique field map. In our analysis of the magnetic stresses, we use the ZDI map corresponding to the inversion using general harmonic field parameterisation (model 2) in \cite{2016A&A...586A..30K}. This magnetic field structure is in agreement with the magnetic map of $\tau$ Sco obtained by \cite{2006MNRAS.370..629D} from a smaller data set.

By fitting theoretical isochrones to luminosity and temperature determined from V mag and B-V color, \cite{2011MNRAS.410..190T} found $\tau$ Sco to have mass $15.0 \pm 0.1 M_{\odot}$ and age  $5.7 \pm 1.0$ Myr. Based on Str\"{o}mgren photometry and Hipparcos parallax, \cite{2012ApJ...746..154P} determine log $T_{eff} = 4.475 \pm 0.073$ and log $L/L_{\odot} = 4.31 \pm 0.16$. They then determine the mass and age from theoretical isochrones to be $14.5 M_{\odot}$ and 5 Myr.  More recently, by using CMFGEN modelling of the spectrum covering wavelength range 380 - 680 nm \citep{2012A&A...538A..29M}, \cite{2021MNRAS.504.2474K} determine $T_{eff} = 31500 \pm 1000$ K, log $g = 4.2 \pm 0.1$ cm s$^{-2}$. From stellar modelling that includes rotational mixing effects and an angular momentum loss rate from equation (\ref{eq:3}), \cite{2021MNRAS.504.2474K} determine from matching $T_{eff}$ and log $g$ that $\tau$ Sco has age less than 6 Myr, but matching the rotation period and nitrogen abundance requires a much larger age, close to that of the terminal age main sequence. To reconcile the observed 41 d rotation period with the predicted rate of spin-down it is found that either the magnetic braking efficiency needs to be larger by a factor of 10 or the initial field was much stronger (30 kG) than the current field. Even then, the observed nitrogen excess requires that rotational mixing be more efficient by at least a factor of 3. From the $T_{eff}$ - log $g$ diagram, the mass is found to be in the range ~ 16 - 18 $M_{\odot}$.  

Based on the Keszthelyi et al. error ellipse in the $T_{eff}$ - log $g$ plane, we find from our DEUCES models that fits can be found for masses between 14.5 and 17.2 $M_{\odot}$. The maximum age is 4.4 Myr. 

There have been a number of age determinations for the Upper Sco association (USco). The general trend is that ages determined from the more massive members are larger by a factor of about 2 than ages from the lower mass stars \cite[see][for discussion and a possible resolution of the discrepancy]{2017ApJ...834...67M}. Since $\tau$ Sco is a B star, here we focus on the USco age determinations based on observations of massive stars. From isochronal ages for the USco B, A, and G stars, as well as the M supergiant Antares, \cite{2012ApJ...746..154P} determined a mean age of $11 \pm 3$ Myr. \citep{2016ApJ...817..164R} determined the age and component masses for seven G- to M-type binary systems in USco using the orbital solutions and HST multi-band photometry. They find that their G-type binaries have ages of ~11.5 Myr, consistent with the age estimate of \cite{2012ApJ...746..154P}. 

Based on its position in the Hertzsprung–Russell diagram, \cite{2014A&A...566A...7N} concluded that $\tau$ Sco is a blue straggler star much younger than its stellar association, which suggests that it formed by stellar merger, a scenario consistent with the suggestion by \cite{2009MNRAS.400L..71F} that fossil magnetic fields originate in stellar merger events, and explored for the specific case of $\tau$ Sco by \cite {2019Natur.574..211S}. The blue straggler interpretation of the $\tau$ Sco data is consistent with the Keszthelyi et al. age determination.

Using the Keszthelyi et al. stellar parameters, we find the mass loss rate is $9.9 \times 10^{-9} M_{\odot}$ yr$^{-1}$, ${v}_{\infty}$ = 2800 km s$^{-1}$ and the solid body rotational inertia is $I = 0.10 MR^2 = 4.8 \times 10^{56}$ g cm$^{2}$.

The ZDI data give  $<B_{eq}^{2}>$ = 0.0194 kG$^{2}$. The spin-down torque, $\tau_{mbw}$, is then $5.6 \times 10^{35}$ dyn cm, which is three orders of magnitude less than found from integration of the ZDI map (see table 1).

Rotational period changes have not been observed for $\tau$ Sco. However, since it is currently a slow rotator, the spin-down time scale is likely to be less than its age, which leads to tension with the spin-down time scale of $t_{mbw} = 4.8 \times 10^{7}$ yr. On the other hand, the spin-down time scale obtained from the ZDI map, $t_{ZDI}$ = $6.6 \times 10^{4}$ yr, is roughly consistent with the current period provided the torque remains roughly constant. 

Fig. 4 is the same as Fig. 1 except now for $\tau$ Sco. Even though the central panel shows a more complicated structure than found for $\sigma$ Ori E, the left panel shows that there are nearly symmetric contributions to the total torque from the Northern and Southern Hemispheres. Unlike $\sigma$ Ori E, there are longitudinal bands in which the contributions to the net torque are positive, in addition to an equatorial band that gives the dominant negative contribution to the net torque.
 
\begin{figure}
	\includegraphics[width=\columnwidth]{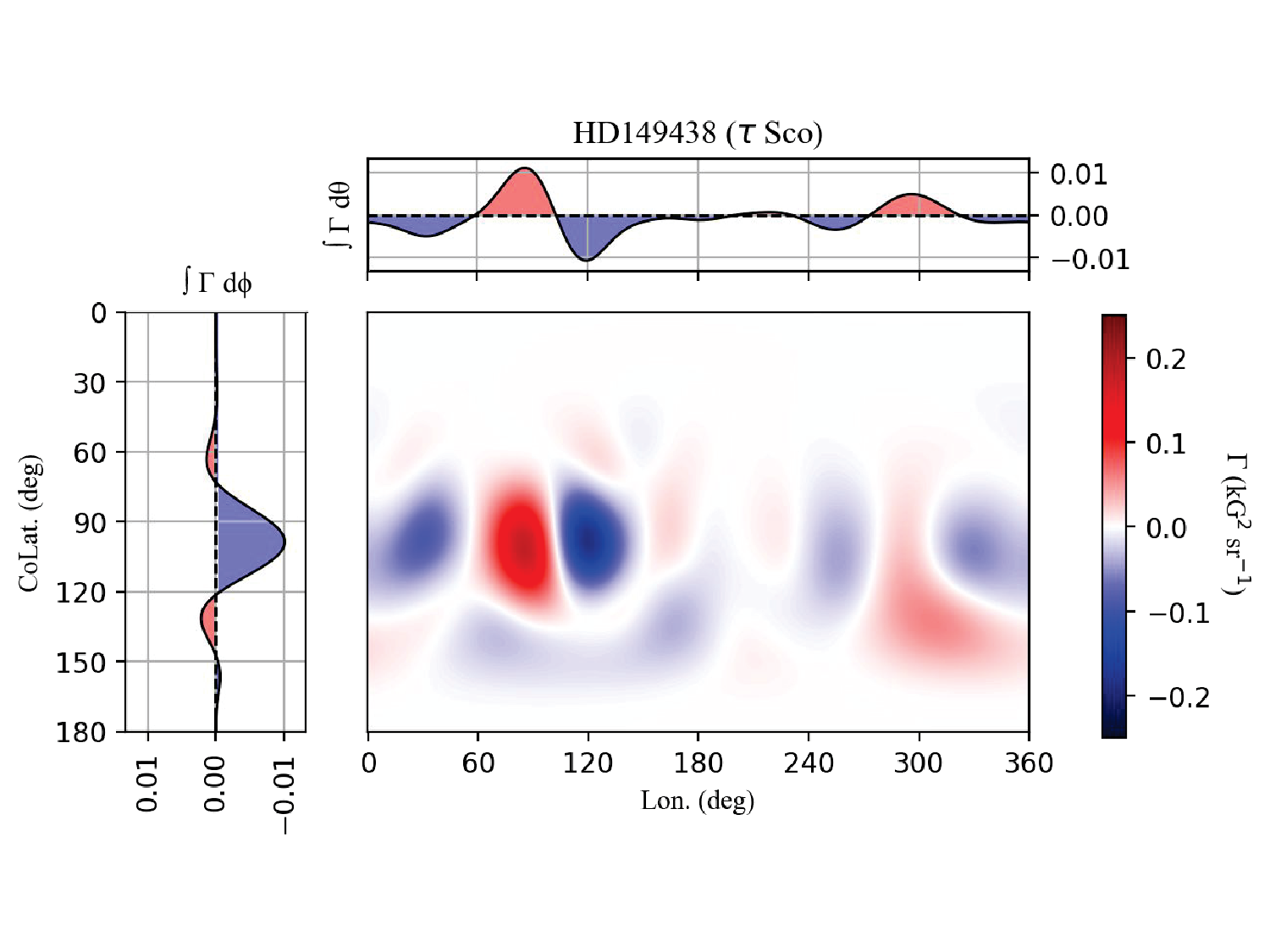}
    \caption{color plot of $\Gamma$ for $\tau$ Sco. $B_{r}$ and $B_{\phi}$ are both in units of kG.}
    \label{fig:figure4}
\end{figure}
 
\subsection{36 Lyn}

36 Lyncis (HD79158) is a helium-weak magnetic Bp star of spectral type B8 III that rotates with period 3.835 d \citep{2006A&A...458..569W}. The field geometry is simpler than that of CU Vir being mainly dipolar but with a strong toroidal component \citep{2018MNRAS.473.3367O}. From fitting optical and UV spectrophotometry, Wade et al. determine luminosity and temperature log $L/L_{\odot} = 2.54 \pm 0.16$, $T_{eff} = 13,300 \pm 300$ K. Based on the solar composition evolutionary calculations of \cite{1992A&AS...96..269S}, Wade et al. determine the stellar mass $M = 4.0 \pm 0.2M_{\odot}$ and age t = 79 - 110 Myr. From our solar composition DEUCES models, if the star is on the main sequence, we find $M = 4.0 \pm 0.3M_{\odot}$ and age t = 110 - 130 Myr. The mass loss rate is $1.4 \times 10^{-11} M_{\odot}$ yr$^{-1}$, and the wind terminal velocity is ${v}_{\infty}$ = 860 km s$^{-1}$. The solid body rotational inertia is $I = 0.059 MR^2 = 2.9 \times 10^{55}$ g cm$^{2}$. If it has evolved just beyond the end of the main sequence, then $M = 3.9 \pm 0.1 M_{\odot}$ and age t = 144 - 164 Myr, the mass loss rate is $\simeq 2 \times 10^{-11} M_{\odot}$ yr$^{-1}$, ${v}_{\infty}$ = 790 km s$^{-1}$and the solid body rotational inertia is  $I \simeq 0.05 MR^2 = 3.0 \times 10^{55}$ g cm$^{2}$.

The ZDI data give $<B_{eq}^{2}>$ = 1.191 kG$^{2}$. The spin-down torque $\tau_{mbw}$, is then between  $6.3 \times 10^{35}$ dyn cm and $1,.3 \times 10^{36}$ dyn cm, which is three orders of magnitude less than found from integration of the ZDI map (see table 1).

Fig. 5 is the same as Fig. 1 except now for 36 Lyn. The central panel shows that different faces of 36 Lyn are dominated by positive and negative contributions to the torque integral. Nevertheless, the left panel reveals symmetry along the rotational equator.

\begin{figure}
	\includegraphics[width=\columnwidth]{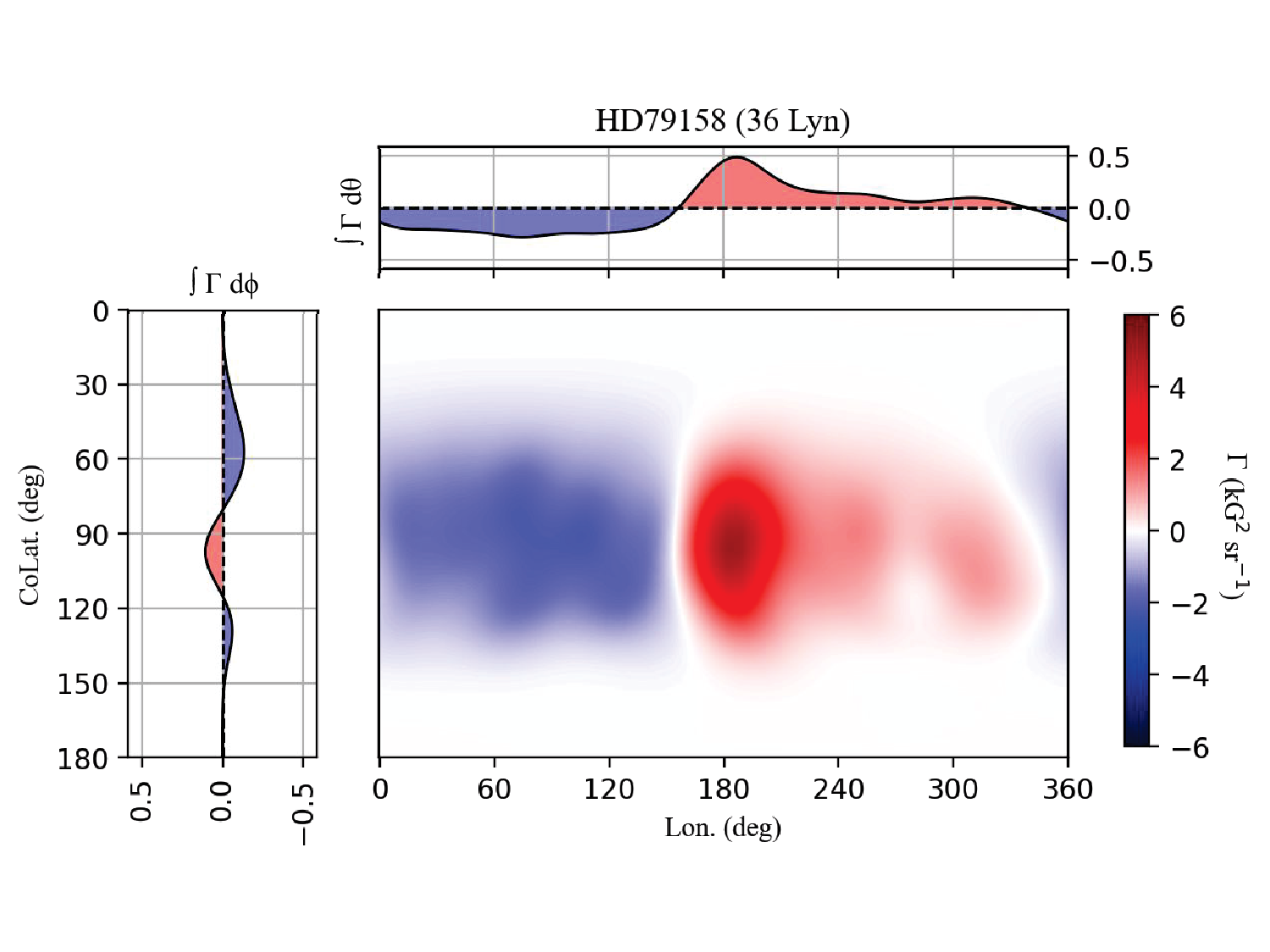}
    \caption{color plot of $\Gamma$ for 36 Lyn. $B_{r}$ and $B_{\phi}$ are both in units of kG.}
    \label{fig:figure5}
\end{figure}

\section{Discussion}

For the fast rotators $\sigma$ Ori E and CU Vir, we find that our estimates of the torque on the stellar surface from ZDI measurements are many orders of magnitude larger than estimated from observed period changes. We stress here that the ZDI measurements only provide the magnetic contribution to the total torque. Since these stars are rapidly rotating and have strong surface magnetic fields deviations from spherical symmetry are likely significant, and as a a consequence the torque due to external pressure (whether thermal or magnetic) acting on the surface will not necessarily be zero.  If the pressure tensor is isotropic, equation (\ref{eq:9}) becomes
 
\begin{equation} \label{eq:18}
 \mathbf{\tau}=-\int\limits_{S}{\left( p+\frac{{{B}^{2}}}{8\pi } \right)\left( \mathbf{r}\times \mathbf{n} \right)dS}+\frac{1}{4\pi }\int\limits_{S}{\left( \mathbf{r}\times \mathbf{B} \right)\left( \mathbf{n}\cdot \mathbf{B} \right)dS}.
\end{equation}

If there are no surface flows, then the total pressure will be constant over the surface and the first term on the right hand side of equation (\ref{eq:18}) will be zero. We can estimate an upper limit on this term by considering the turbulent velocity derived from spectroscopic measurements. \cite{2018MNRAS.475.5144S} find for $\sigma$ Ori E, a macroturbulence velocity $v_{mac} =13 \pm 16$ km s$^{-1}$ across 7 different lines. Most of the lines are consistent with zero turbulent broadening; the 13 km s$^{-1}$ value is the result of a couple of lines giving higher values, which probably are outliers. If we conservatively adopt an upper limit of 30 km s$^{-1}$ for the velocity of any surface flows, using the photospheric density from our stellar model, $\rho = 1.3 \times 10^{-9}$ g cm$^{-3}$, we find an upper limit of $~  10^{39}$ dyn cm for the first term on the right hand side of equation (\ref{eq:18}), which is small compared to the magnitude of the torque,  $7 \times 10^{40}$ dyn cm, derived from magnetic stresses. Also $\sigma$ Ori E, 36 Lyn and CU Vir have prominent, high-contrast chemical abundance spots on their surfaces. These spots are stable, as can be judged from the lack of changes in the light curve shapes \citep{2013ApJ...769...33T}. The existence of these spots is likely incompatible with significant surface flows.

Our models indicate that the thermal pressure at the photosphere of $\sigma$ Ori E, $~ 5 \times 10^3$ dyn cm$^{-2}$, is negligible compared to the magnetic pressure, $~ 2 \times 10^7$ dyn cm$^{-2}$. Indeed all of the stars considered here have magnetically dominated photospheres, with $\beta$ in the range $2 \times 10^{-4}$ for $\sigma$ Ori E up to order unity for $\tau$ Sco.

Since it has been found for cool stars that ZDI mapping is insensitive to small scale magnetic fields \citep{2010MNRAS.404..101J}, they may also play a role in determining the magnetic force and torque. However, in the case of magnetic CP stars, the notion of the dichotomy between large/small-scale fields and ZDI/MI modelling methodologies is known to be not applicable. For mCP stars, the treatment of Zeeman broadening/intensification has always been an integral part of ZDI, particularly when individual lines were modelled \citep{2015A&A...574A..79K, 2018A&A...609A..88R}. Even before ZDI, global field models were capable of fitting phase curves of longitudinal field and mean field modulus with "pure" large-scale field geometries without the need of invoking hidden small-scale field \citep[e.g.][] {2000A&A...359..213L, 2002A&A...394.1023B}. These studies, and direct observations of well-resolved Zeeman split lines \citep[e.g.][] {1997A&AS..123..353M, 2017A&A...601A..14M}, indicate that magnetic CP stars possess no small-scale fields of the type existing in cool stars. We have analyzed how fields at different scales contribute to the spindown torque (see figures A5 - A8, and the associated text) and found that the resulting torques were mainly determined by large and medium scale fields.

Re-accretion of wind material that gets trapped by the magnetic field could provide a torque in the direction of increasing the rotation rate of the star, i.e. in the opposite sense found here from the ZDI maps. However, the fast rotators are in the centrifugally dominated regime \citep{2013MNRAS.429..398P} and material that breaks free from the magnetic field will mainly be ejected and not accreted \citep{2020MNRAS.499.5366O}. A centrifugal breakout episode may have been recently observed for CU Vir \citep{2021ApJ...921....9D}.

Due to the large non-uniform magnetic pressure and the insignificance of thermal and turbulence pressure, it seems unavoidable that the stellar surface will deviate from the spherical symmetry assumed in our force and torque calculations. Indeed for the fastest rotator, CU Vir, we estimate by assuming the photosphere is an equipotential surface, that centrifugal effects lead to an equatorial to polar radius ratio of ~ 1.07. The possibility arises that asphericity is responsible for our finding that in some cases the torques calculated from integration of the Maxwell stresses over the stellar surface are significantly larger than found from the magnetically braked wind analysis.

\section{Conclusions}

We have used the distributions of the surface magnetic field components determined from ZDI measurements to calculate the net force and torque resulting from magnetic stresses for 3 rapidly rotating magnetic chemically peculiar stars, CU Vir, 36 Lyn, and $\sigma$ Ori E, and the slowly rotating early B-type star $\tau$ Sco. We find that if we assume solid body rotation, the spin down times are up to four orders of magnitude shorter than the stellar age. If the magnetic fields are truly fossil in nature, then they are frozen in, and are expected to be tied to the surface radius such that the surface field strength decreases as $1/R^2$. As the torque is proportional to $R^3 B^2$, this means that throughout the stellar evolution, the torques would vary as $1/R$ (getting smaller with time). However, the change in radius is not expected to be significant until e.g. the end of the main sequence. Therefore an assumption of constant torque is a good one. If anything, it would mean that, from a stellar evolution point of view, the torque was slightly larger in the past, whereas we would need the torque to be significantly smaller in the past to reconcile the large current spin down timescale obtained by our method and the presumed stellar ages. 

 We have considered a number of ways that this dilemma might be avoided and conclude that the mostly likely resolution is that our assumption of exact spherically symmetry is violated.  Furthermore in all cases, the photospheres are magnetically dominated and the deviations from spherical symmetry will arise from the nonuniform magnetic stresses. For the slowest rotator in the sample, $\tau$ Sco, centrifugal distortion of the photosphere is negligible, and any deviations from sphericity must result from the magnetic stresses.

\section*{Acknowledgements}

This research is supported in part by NSF Grant AST-2108455 (VP and JM), NASA National Space Grant NNX15AI19H (JM), the University of Delaware Annie Jump Cannon Fund (MES), and  the Swedish Research Council grants 2019-03548 and 2023-03667 (OK). VP thanks Aline Vidotto for informative discussions. We thank the anonymous referee for comments that lead to improvement of the manuscript.
\section*{Data Availability}

No new data were generated or analyzed in support of this research.



\bibliographystyle{mnras}
\bibliography{refs} 

\begin{thebibliography}{}
\makeatletter
\relax
\def\mn@urlcharsother{\let\do\@makeother \do\$\do\&\do\#\do\^\do\_\do\%\do\~}
\def\mn@doi{\begingroup\mn@urlcharsother \@ifnextchar [ {\mn@doi@}
  {\mn@doi@[]}}
\def\mn@doi@[#1]#2{\def\@tempa{#1}\ifx\@tempa\@empty \href
  {http://dx.doi.org/#2} {doi:#2}\else \href {http://dx.doi.org/#2} {#1}\fi
  \endgroup}
\def\mn@eprint#1#2{\mn@eprint@#1:#2::\@nil}
\def\mn@eprint@arXiv#1{\href {http://arxiv.org/abs/#1} {{\tt arXiv:#1}}}
\def\mn@eprint@dblp#1{\href {http://dblp.uni-trier.de/rec/bibtex/#1.xml}
  {dblp:#1}}
\def\mn@eprint@#1:#2:#3:#4\@nil{\def\@tempa {#1}\def\@tempb {#2}\def\@tempc
  {#3}\ifx \@tempc \@empty \let \@tempc \@tempb \let \@tempb \@tempa \fi \ifx
  \@tempb \@empty \def\@tempb {arXiv}\fi \@ifundefined
  {mn@eprint@\@tempb}{\@tempb:\@tempc}{\expandafter \expandafter \csname
  mn@eprint@\@tempb\endcsname \expandafter{\@tempc}}}

\bibitem[\protect\citeauthoryear{{Abt} \& {Morrell}}{{Abt} \&
  {Morrell}}{1995}]{1995ApJS...99..135A}
{Abt} H.~A.,  {Morrell} N.~I.,  1995, \mn@doi [\apjs] {10.1086/192182}, \href
  {https://ui.adsabs.harvard.edu/abs/1995ApJS...99..135A} {99, 135}

\bibitem[\protect\citeauthoryear{{Bagnulo}, {Landi Degl'Innocenti}, {Landolfi}
  \& {Mathys}}{{Bagnulo} et~al.}{2002}]{2002A&A...394.1023B}
{Bagnulo} S.,  {Landi Degl'Innocenti} M.,  {Landolfi} M.,   {Mathys} G.,  2002,
  \mn@doi [\aap] {10.1051/0004-6361:20021221}, \href
  {https://ui.adsabs.harvard.edu/abs/2002A&A...394.1023B} {394, 1023}

\bibitem[\protect\citeauthoryear{{Belcher} \& {MacGregor}}{{Belcher} \&
  {MacGregor}}{1976}]{1976ApJ...210..498B}
{Belcher} J.~W.,  {MacGregor} K.~B.,  1976, \mn@doi [\apj] {10.1086/154853},
  \href {https://ui.adsabs.harvard.edu/abs/1976ApJ...210..498B} {210, 498}

\bibitem[\protect\citeauthoryear{{Blanco} \& {Catalano}}{{Blanco} \&
  {Catalano}}{1971}]{1971AJ.....76..630B}
{Blanco} C.,  {Catalano} F.~A.,  1971, \mn@doi [\aj] {10.1086/111175}, \href
  {https://ui.adsabs.harvard.edu/abs/1971AJ.....76..630B} {76, 630}

\bibitem[\protect\citeauthoryear{{Das} \& {Chandra}}{{Das} \&
  {Chandra}}{2021}]{2021ApJ...921....9D}
{Das} B.,  {Chandra} P.,  2021, \mn@doi [\apj] {10.3847/1538-4357/ac1075},
  \href {https://ui.adsabs.harvard.edu/abs/2021ApJ...921....9D} {921, 9}

\bibitem[\protect\citeauthoryear{{Donati} \& {Landstreet}}{{Donati} \&
  {Landstreet}}{2009}]{2009ARA&A..47..333D}
{Donati} J.~F.,  {Landstreet} J.~D.,  2009, \mn@doi [\araa]
  {10.1146/annurev-astro-082708-101833}, \href
  {https://ui.adsabs.harvard.edu/abs/2009ARA&A..47..333D} {47, 333}

\bibitem[\protect\citeauthoryear{{Donati} et~al.,}{{Donati}
  et~al.}{2006}]{2006MNRAS.370..629D}
{Donati} J.~F.,  et~al., 2006, \mn@doi [\mnras]
  {10.1111/j.1365-2966.2006.10558.x}, \href
  {https://ui.adsabs.harvard.edu/abs/2006MNRAS.370..629D} {370, 629}

\bibitem[\protect\citeauthoryear{{Erba}, {Shultz}, {Petit}, {Fullerton},
  {Henrichs}, {Kochukhov}, {Rivinius}  \& {Wade}}{{Erba}
  et~al.}{2021}]{2021MNRAS.506.2296E}
{Erba} C.,  {Shultz} M.~E.,  {Petit} V.,  {Fullerton} A.~W.,  {Henrichs} H.~F.,
   {Kochukhov} O.,  {Rivinius} T.,   {Wade} G.~A.,  2021, \mn@doi [\mnras]
  {10.1093/mnras/stab1454}, \href
  {https://ui.adsabs.harvard.edu/abs/2021MNRAS.506.2296E} {506, 2296}

\bibitem[\protect\citeauthoryear{{Ferrario}, {Pringle}, {Tout}  \&
  {Wickramasinghe}}{{Ferrario} et~al.}{2009}]{2009MNRAS.400L..71F}
{Ferrario} L.,  {Pringle} J.~E.,  {Tout} C.~A.,   {Wickramasinghe} D.~T.,
  2009, \mn@doi [\mnras] {10.1111/j.1745-3933.2009.00765.x}, \href
  {https://ui.adsabs.harvard.edu/abs/2009MNRAS.400L..71F} {400, L71}

\bibitem[\protect\citeauthoryear{{Groote} \& {Hunger}}{{Groote} \&
  {Hunger}}{1982}]{1982A&A...116...64G}
{Groote} D.,  {Hunger} K.,  1982, \aap, \href
  {https://ui.adsabs.harvard.edu/abs/1982A&A...116...64G} {116, 64}

\bibitem[\protect\citeauthoryear{{Grunhut} et~al.,}{{Grunhut}
  et~al.}{2012}]{2012MNRAS.419.1610G}
{Grunhut} J.~H.,  et~al., 2012, \mn@doi [\mnras]
  {10.1111/j.1365-2966.2011.19824.x}, \href
  {https://ui.adsabs.harvard.edu/abs/2012MNRAS.419.1610G} {419, 1610}

\bibitem[\protect\citeauthoryear{{Grunhut} et~al.,}{{Grunhut}
  et~al.}{2017}]{2017MNRAS.465.2432G}
{Grunhut} J.~H.,  et~al., 2017, \mn@doi [\mnras] {10.1093/mnras/stw2743}, \href
  {https://ui.adsabs.harvard.edu/abs/2017MNRAS.465.2432G} {465, 2432}

\bibitem[\protect\citeauthoryear{{Johnstone}, {Jardine}  \&
  {Mackay}}{{Johnstone} et~al.}{2010}]{2010MNRAS.404..101J}
{Johnstone} C.,  {Jardine} M.,   {Mackay} D.~H.,  2010, \mn@doi [\mnras]
  {10.1111/j.1365-2966.2010.16298.x}, \href
  {https://ui.adsabs.harvard.edu/abs/2010MNRAS.404..101J} {404, 101}

\bibitem[\protect\citeauthoryear{{Kawaler}}{{Kawaler}}{1988}]{1988ApJ...333..236K}
{Kawaler} S.~D.,  1988, \mn@doi [\apj] {10.1086/166740}, \href
  {https://ui.adsabs.harvard.edu/abs/1988ApJ...333..236K} {333, 236}

\bibitem[\protect\citeauthoryear{{Keszthelyi}, {Meynet}, {Martins}, {de Koter}
  \& {David-Uraz}}{{Keszthelyi} et~al.}{2021}]{2021MNRAS.504.2474K}
{Keszthelyi} Z.,  {Meynet} G.,  {Martins} F.,  {de Koter} A.,   {David-Uraz}
  A.,  2021, \mn@doi [\mnras] {10.1093/mnras/stab893}, \href
  {https://ui.adsabs.harvard.edu/abs/2021MNRAS.504.2474K} {504, 2474}

\bibitem[\protect\citeauthoryear{{Kochukhov} \& {Wade}}{{Kochukhov} \&
  {Wade}}{2016}]{2016A&A...586A..30K}
{Kochukhov} O.,  {Wade} G.~A.,  2016, \mn@doi [\aap]
  {10.1051/0004-6361/201527454}, \href
  {https://ui.adsabs.harvard.edu/abs/2016A&A...586A..30K} {586, A30}

\bibitem[\protect\citeauthoryear{{Kochukhov}, {L{\"u}ftinger}, {Neiner},
  {Alecian}  \& {MiMeS Collaboration}}{{Kochukhov}
  et~al.}{2014}]{2014A&A...565A..83K}
{Kochukhov} O.,  {L{\"u}ftinger} T.,  {Neiner} C.,  {Alecian} E.,   {MiMeS
  Collaboration} 2014, \mn@doi [\aap] {10.1051/0004-6361/201423472}, \href
  {https://ui.adsabs.harvard.edu/abs/2014A&A...565A..83K} {565, A83}

\bibitem[\protect\citeauthoryear{{Kochukhov} et~al.,}{{Kochukhov}
  et~al.}{2015}]{2015A&A...574A..79K}
{Kochukhov} O.,  et~al., 2015, \mn@doi [\aap] {10.1051/0004-6361/201425065},
  \href {https://ui.adsabs.harvard.edu/abs/2015A&A...574A..79K} {574, A79}

\bibitem[\protect\citeauthoryear{{Krti{\v{c}}ka} et~al.,}{{Krti{\v{c}}ka}
  et~al.}{2019}]{2019A&A...625A..34K}
{Krti{\v{c}}ka} J.,  et~al., 2019, \mn@doi [\aap]
  {10.1051/0004-6361/201834937}, \href
  {https://ui.adsabs.harvard.edu/abs/2019A&A...625A..34K} {625, A34}

\bibitem[\protect\citeauthoryear{{Kuschnig}, {Ryabchikova}, {Piskunov}, {Weiss}
   \& {Gelbmann}}{{Kuschnig} et~al.}{1999}]{1999A&A...348..924K}
{Kuschnig} R.,  {Ryabchikova} T.~A.,  {Piskunov} N.~E.,  {Weiss} W.~W.,
  {Gelbmann} M.~J.,  1999, \aap, \href
  {https://ui.adsabs.harvard.edu/abs/1999A&A...348..924K} {348, 924}

\bibitem[\protect\citeauthoryear{{Landstreet} \& {Mathys}}{{Landstreet} \&
  {Mathys}}{2000}]{2000A&A...359..213L}
{Landstreet} J.~D.,  {Mathys} G.,  2000, \aap, \href
  {https://ui.adsabs.harvard.edu/abs/2000A&A...359..213L} {359, 213}

\bibitem[\protect\citeauthoryear{{Lawlor} \& {MacDonald}}{{Lawlor} \&
  {MacDonald}}{2023}]{2023MNRAS.525.4700L}
{Lawlor} T.~M.,  {MacDonald} J.,  2023, \mn@doi [\mnras]
  {10.1093/mnras/stad2582}, \href
  {https://ui.adsabs.harvard.edu/abs/2023MNRAS.525.4700L} {525, 4700}

\bibitem[\protect\citeauthoryear{{Lipski} \& {St{\c{e}}pie{\'n}}}{{Lipski} \&
  {St{\c{e}}pie{\'n}}}{2008}]{2008MNRAS.385..481L}
{Lipski} {\L}.,  {St{\c{e}}pie{\'n}} K.,  2008, \mn@doi [\mnras]
  {10.1111/j.1365-2966.2008.12856.x}, \href
  {https://ui.adsabs.harvard.edu/abs/2008MNRAS.385..481L} {385, 481}

\bibitem[\protect\citeauthoryear{{MacDonald} \& {Mullan}}{{MacDonald} \&
  {Mullan}}{2017}]{2017ApJ...834...67M}
{MacDonald} J.,  {Mullan} D.~J.,  2017, \mn@doi [\apj]
  {10.3847/1538-4357/834/1/67}, \href
  {https://ui.adsabs.harvard.edu/abs/2017ApJ...834...67M} {834, 67}

\bibitem[\protect\citeauthoryear{{Martins}, {Escolano}, {Wade}, {Donati},
  {Bouret}  \& {Mimes Collaboration}}{{Martins}
  et~al.}{2012}]{2012A&A...538A..29M}
{Martins} F.,  {Escolano} C.,  {Wade} G.~A.,  {Donati} J.~F.,  {Bouret} J.~C.,
   {Mimes Collaboration} 2012, \mn@doi [\aap] {10.1051/0004-6361/201118039},
  \href {https://ui.adsabs.harvard.edu/abs/2012A&A...538A..29M} {538, A29}

\bibitem[\protect\citeauthoryear{{Mathys}}{{Mathys}}{2017}]{2017A&A...601A..14M}
{Mathys} G.,  2017, \mn@doi [\aap] {10.1051/0004-6361/201628429}, \href
  {https://ui.adsabs.harvard.edu/abs/2017A&A...601A..14M} {601, A14}

\bibitem[\protect\citeauthoryear{{Mathys}, {Hubrig}, {Landstreet}, {Lanz}  \&
  {Manfroid}}{{Mathys} et~al.}{1997}]{1997A&AS..123..353M}
{Mathys} G.,  {Hubrig} S.,  {Landstreet} J.~D.,  {Lanz} T.,   {Manfroid} J.,
  1997, \mn@doi [\aaps] {10.1051/aas:1997103}, \href
  {https://ui.adsabs.harvard.edu/abs/1997A&AS..123..353M} {123, 353}

\bibitem[\protect\citeauthoryear{{Matt} \& {Pudritz}}{{Matt} \&
  {Pudritz}}{2008}]{2008ApJ...678.1109M}
{Matt} S.,  {Pudritz} R.~E.,  2008, \mn@doi [\apj] {10.1086/533428}, \href
  {https://ui.adsabs.harvard.edu/abs/2008ApJ...678.1109M} {678, 1109}

\bibitem[\protect\citeauthoryear{{Matt}, {MacGregor}, {Pinsonneault}  \&
  {Greene}}{{Matt} et~al.}{2012}]{2012ApJ...754L..26M}
{Matt} S.~P.,  {MacGregor} K.~B.,  {Pinsonneault} M.~H.,   {Greene} T.~P.,
  2012, \mn@doi [\apjl] {10.1088/2041-8205/754/2/L26}, \href
  {https://ui.adsabs.harvard.edu/abs/2012ApJ...754L..26M} {754, L26}

\bibitem[\protect\citeauthoryear{{Mestel}}{{Mestel}}{1968}]{1968MNRAS.138..359M}
{Mestel} L.,  1968, \mn@doi [\mnras] {10.1093/mnras/138.3.359}, \href
  {https://ui.adsabs.harvard.edu/abs/1968MNRAS.138..359M} {138, 359}

\bibitem[\protect\citeauthoryear{{Mestel}}{{Mestel}}{1984}]{1984LNP...193...49M}
{Mestel} L.,  1984, in {Baliunas} S.~L.,  {Hartmann} L.,  eds, , Vol.~193, Cool
  Stars, Stellar Systems, and the Sun.
p.~49, \mn@doi{10.1007/3-540-12907-3\_179}

\bibitem[\protect\citeauthoryear{{Mestel} \& {Selley}}{{Mestel} \&
  {Selley}}{1970}]{1970MNRAS.149..197M}
{Mestel} L.,  {Selley} C.~S.,  1970, \mn@doi [\mnras]
  {10.1093/mnras/149.3.197}, \href
  {https://ui.adsabs.harvard.edu/abs/1970MNRAS.149..197M} {149, 197}

\bibitem[\protect\citeauthoryear{{Mestel} \& {Spruit}}{{Mestel} \&
  {Spruit}}{1987}]{1987MNRAS.226...57M}
{Mestel} L.,  {Spruit} H.~C.,  1987, \mn@doi [\mnras] {10.1093/mnras/226.1.57},
  \href {https://ui.adsabs.harvard.edu/abs/1987MNRAS.226...57M} {226, 57}

\bibitem[\protect\citeauthoryear{{Mikul{\'a}{\v{s}}ek}
  et~al.,}{{Mikul{\'a}{\v{s}}ek} et~al.}{2008}]{2008A&A...485..585M}
{Mikul{\'a}{\v{s}}ek} Z.,  et~al., 2008, \mn@doi [\aap]
  {10.1051/0004-6361:20077794}, \href
  {https://ui.adsabs.harvard.edu/abs/2008A&A...485..585M} {485, 585}

\bibitem[\protect\citeauthoryear{{Mikul{\'a}{\v{s}}ek}
  et~al.,}{{Mikul{\'a}{\v{s}}ek} et~al.}{2011}]{2011A&A...534L...5M}
{Mikul{\'a}{\v{s}}ek} Z.,  et~al., 2011, \mn@doi [\aap]
  {10.1051/0004-6361/201117784}, \href
  {https://ui.adsabs.harvard.edu/abs/2011A&A...534L...5M} {534, L5}

\bibitem[\protect\citeauthoryear{{Mikul{\'a}{\v{s}}ek}, {Krti{\v{c}}ka},
  {Jan{\'\i}k}, {Henry}, {Zejda}, {Shultz}, {Paunzen}  \&
  {Jagelka}}{{Mikul{\'a}{\v{s}}ek} et~al.}{2017}]{2017ASPC..510..220M}
{Mikul{\'a}{\v{s}}ek} Z.~Z.,  {Krti{\v{c}}ka} J.,  {Jan{\'\i}k} J.,  {Henry}
  G.~W.,  {Zejda} M.,  {Shultz} M.,  {Paunzen} E.,   {Jagelka} M.,  2017, in
  {Balega} Y.~Y.,  {Kudryavtsev} D.~O.,  {Romanyuk} I.~I.,   {Yakunin} I.~A.,
  eds,  Astronomical Society of the Pacific Conference Series Vol. 510, Stars:
  From Collapse to Collapse. p.~220 (\mn@eprint {arXiv} {1702.02195})

\bibitem[\protect\citeauthoryear{{Nielsen}, {Gizon}, {Schunker}  \&
  {Karoff}}{{Nielsen} et~al.}{2013}]{2013A&A...557L..10N}
{Nielsen} M.~B.,  {Gizon} L.,  {Schunker} H.,   {Karoff} C.,  2013, \mn@doi
  [\aap] {10.1051/0004-6361/201321912}, \href
  {https://ui.adsabs.harvard.edu/abs/2013A&A...557L..10N} {557, L10}

\bibitem[\protect\citeauthoryear{{Nieva} \& {Przybilla}}{{Nieva} \&
  {Przybilla}}{2014}]{2014A&A...566A...7N}
{Nieva} M.-F.,  {Przybilla} N.,  2014, \mn@doi [\aap]
  {10.1051/0004-6361/201423373}, \href
  {https://ui.adsabs.harvard.edu/abs/2014A&A...566A...7N} {566, A7}

\bibitem[\protect\citeauthoryear{{Oksala} et~al.,}{{Oksala}
  et~al.}{2015}]{2015MNRAS.451.2015O}
{Oksala} M.~E.,  et~al., 2015, \mn@doi [\mnras] {10.1093/mnras/stv1086}, \href
  {https://ui.adsabs.harvard.edu/abs/2015MNRAS.451.2015O} {451, 2015}

\bibitem[\protect\citeauthoryear{{Oksala}, {Silvester}, {Kochukhov}, {Neiner},
  {Wade}  \& {MiMeS Collaboration}}{{Oksala}
  et~al.}{2018}]{2018MNRAS.473.3367O}
{Oksala} M.~E.,  {Silvester} J.,  {Kochukhov} O.,  {Neiner} C.,  {Wade} G.~A.,
   {MiMeS Collaboration} 2018, \mn@doi [\mnras] {10.1093/mnras/stx2487}, \href
  {https://ui.adsabs.harvard.edu/abs/2018MNRAS.473.3367O} {473, 3367}

\bibitem[\protect\citeauthoryear{{Owocki}, {Shultz}, {ud-Doula}, {Sundqvist},
  {Townsend}  \& {Cranmer}}{{Owocki} et~al.}{2020}]{2020MNRAS.499.5366O}
{Owocki} S.~P.,  {Shultz} M.~E.,  {ud-Doula} A.,  {Sundqvist} J.~O.,
  {Townsend} R. H.~D.,   {Cranmer} S.~R.,  2020, \mn@doi [\mnras]
  {10.1093/mnras/staa2325}, \href
  {https://ui.adsabs.harvard.edu/abs/2020MNRAS.499.5366O} {499, 5366}

\bibitem[\protect\citeauthoryear{{Parker}}{{Parker}}{1969}]{1969SSRv....9..325P}
{Parker} E.~N.,  1969, \mn@doi [\ssr] {10.1007/BF00175236}, \href
  {https://ui.adsabs.harvard.edu/abs/1969SSRv....9..325P} {9, 325}

\bibitem[\protect\citeauthoryear{{Pecaut}, {Mamajek}  \& {Bubar}}{{Pecaut}
  et~al.}{2012}]{2012ApJ...746..154P}
{Pecaut} M.~J.,  {Mamajek} E.~E.,   {Bubar} E.~J.,  2012, \mn@doi [\apj]
  {10.1088/0004-637X/746/2/154}, \href
  {https://ui.adsabs.harvard.edu/abs/2012ApJ...746..154P} {746, 154}

\bibitem[\protect\citeauthoryear{{Petit} et~al.,}{{Petit}
  et~al.}{2013}]{2013MNRAS.429..398P}
{Petit} V.,  et~al., 2013, \mn@doi [\mnras] {10.1093/mnras/sts344}, \href
  {https://ui.adsabs.harvard.edu/abs/2013MNRAS.429..398P} {429, 398}

\bibitem[\protect\citeauthoryear{{Pyper} \& {Adelman}}{{Pyper} \&
  {Adelman}}{2020}]{2020PASP..132b4201P}
{Pyper} D.~M.,  {Adelman} S.~J.,  2020, \mn@doi [\pasp]
  {10.1088/1538-3873/ab6279}, \href
  {https://ui.adsabs.harvard.edu/abs/2020PASP..132b4201P} {132, 024201}

\bibitem[\protect\citeauthoryear{{Pyper}, {Ryabchikova}, {Malanushenko},
  {Kuschnig}, {Plachinda}  \& {Savanov}}{{Pyper}
  et~al.}{1998}]{1998A&A...339..822P}
{Pyper} D.~M.,  {Ryabchikova} T.,  {Malanushenko} V.,  {Kuschnig} R.,
  {Plachinda} S.,   {Savanov} I.,  1998, \aap, \href
  {https://ui.adsabs.harvard.edu/abs/1998A&A...339..822P} {339, 822}

\bibitem[\protect\citeauthoryear{{Pyper}, {Stevens}  \& {Adelman}}{{Pyper}
  et~al.}{2013}]{2013MNRAS.431.2106P}
{Pyper} D.~M.,  {Stevens} I.~R.,   {Adelman} S.~J.,  2013, \mn@doi [\mnras]
  {10.1093/mnras/stt256}, \href
  {https://ui.adsabs.harvard.edu/abs/2013MNRAS.431.2106P} {431, 2106}

\bibitem[\protect\citeauthoryear{{Reiners} \& {Mohanty}}{{Reiners} \&
  {Mohanty}}{2012}]{2012ApJ...746...43R}
{Reiners} A.,  {Mohanty} S.,  2012, \mn@doi [\apj]
  {10.1088/0004-637X/746/1/43}, \href
  {https://ui.adsabs.harvard.edu/abs/2012ApJ...746...43R} {746, 43}

\bibitem[\protect\citeauthoryear{{R{\'e}ville}, {Brun}, {Matt}, {Strugarek}  \&
  {Pinto}}{{R{\'e}ville} et~al.}{2015}]{2015ApJ...798..116R}
{R{\'e}ville} V.,  {Brun} A.~S.,  {Matt} S.~P.,  {Strugarek} A.,   {Pinto}
  R.~F.,  2015, \mn@doi [\apj] {10.1088/0004-637X/798/2/116}, \href
  {https://ui.adsabs.harvard.edu/abs/2015ApJ...798..116R} {798, 116}

\bibitem[\protect\citeauthoryear{{Rivinius}, {Townsend}, {Kochukhov},
  {{\v{S}}tefl}, {Baade}, {Barrera}  \& {Szeifert}}{{Rivinius}
  et~al.}{2013}]{2013MNRAS.429..177R}
{Rivinius} T.,  {Townsend} R.~H.~D.,  {Kochukhov} O.,  {{\v{S}}tefl} S.,
  {Baade} D.,  {Barrera} L.,   {Szeifert} T.,  2013, \mn@doi [\mnras]
  {10.1093/mnras/sts323}, \href
  {https://ui.adsabs.harvard.edu/abs/2013MNRAS.429..177R} {429, 177}

\bibitem[\protect\citeauthoryear{{Rizzuto}, {Ireland}, {Dupuy}  \&
  {Kraus}}{{Rizzuto} et~al.}{2016}]{2016ApJ...817..164R}
{Rizzuto} A.~C.,  {Ireland} M.~J.,  {Dupuy} T.~J.,   {Kraus} A.~L.,  2016,
  \mn@doi [\apj] {10.3847/0004-637X/817/2/164}, \href
  {https://ui.adsabs.harvard.edu/abs/2016ApJ...817..164R} {817, 164}

\bibitem[\protect\citeauthoryear{{Rusomarov}, {Kochukhov}  \&
  {Lundin}}{{Rusomarov} et~al.}{2018}]{2018A&A...609A..88R}
{Rusomarov} N.,  {Kochukhov} O.,   {Lundin} A.,  2018, \mn@doi [\aap]
  {10.1051/0004-6361/201731914}, \href
  {https://ui.adsabs.harvard.edu/abs/2018A&A...609A..88R} {609, A88}

\bibitem[\protect\citeauthoryear{{Sakurai}}{{Sakurai}}{1985}]{1985A&A...152..121S}
{Sakurai} T.,  1985, \aap, \href
  {https://ui.adsabs.harvard.edu/abs/1985A&A...152..121S} {152, 121}

\bibitem[\protect\citeauthoryear{{Schaller}, {Schaerer}, {Meynet}  \&
  {Maeder}}{{Schaller} et~al.}{1992}]{1992A&AS...96..269S}
{Schaller} G.,  {Schaerer} D.,  {Meynet} G.,   {Maeder} A.,  1992, \aaps, \href
  {https://ui.adsabs.harvard.edu/abs/1992A&AS...96..269S} {96, 269}

\bibitem[\protect\citeauthoryear{{Schatzman}}{{Schatzman}}{1962}]{1962AnAp...25...18S}
{Schatzman} E.,  1962, Annales d'Astrophysique, \href
  {https://ui.adsabs.harvard.edu/abs/1962AnAp...25...18S} {25, 18}

\bibitem[\protect\citeauthoryear{{Schneider}, {Ohlmann}, {Podsiadlowski},
  {R{\"o}pke}, {Balbus}, {Pakmor}  \& {Springel}}{{Schneider}
  et~al.}{2019}]{2019Natur.574..211S}
{Schneider} F. R.~N.,  {Ohlmann} S.~T.,  {Podsiadlowski} P.,  {R{\"o}pke}
  F.~K.,  {Balbus} S.~A.,  {Pakmor} R.,   {Springel} V.,  2019, \mn@doi [\nat]
  {10.1038/s41586-019-1621-5}, \href
  {https://ui.adsabs.harvard.edu/abs/2019Natur.574..211S} {574, 211}

\bibitem[\protect\citeauthoryear{{Sherry}, {Walter}, {Wolk}  \&
  {Adams}}{{Sherry} et~al.}{2008}]{2008AJ....135.1616S}
{Sherry} W.~H.,  {Walter} F.~M.,  {Wolk} S.~J.,   {Adams} N.~R.,  2008, \mn@doi
  [\aj] {10.1088/0004-6256/135/4/1616}, \href
  {https://ui.adsabs.harvard.edu/abs/2008AJ....135.1616S} {135, 1616}

\bibitem[\protect\citeauthoryear{{Shultz}, {Wade}, {Rivinius}, {Neiner},
  {Henrichs}, {Marcolino}  \& {MiMeS Collaboration}}{{Shultz}
  et~al.}{2017}]{2017MNRAS.471.2286S}
{Shultz} M.,  {Wade} G.~A.,  {Rivinius} T.,  {Neiner} C.,  {Henrichs} H.,
  {Marcolino} W.,   {MiMeS Collaboration} 2017, \mn@doi [\mnras]
  {10.1093/mnras/stx1632}, \href
  {https://ui.adsabs.harvard.edu/abs/2017MNRAS.471.2286S} {471, 2286}

\bibitem[\protect\citeauthoryear{{Shultz} et~al.,}{{Shultz}
  et~al.}{2018}]{2018MNRAS.475.5144S}
{Shultz} M.~E.,  et~al., 2018, \mn@doi [\mnras] {10.1093/mnras/sty103}, \href
  {https://ui.adsabs.harvard.edu/abs/2018MNRAS.475.5144S} {475, 5144}

\bibitem[\protect\citeauthoryear{{Shultz} et~al.,}{{Shultz}
  et~al.}{2019a}]{2019MNRAS.485.1508S}
{Shultz} M.~E.,  et~al., 2019a, \mn@doi [\mnras] {10.1093/mnras/stz416}, \href
  {https://ui.adsabs.harvard.edu/abs/2019MNRAS.485.1508S} {485, 1508}

\bibitem[\protect\citeauthoryear{{Shultz}, {Rivinius}, {Das}, {Wade}  \&
  {Chandra}}{{Shultz} et~al.}{2019b}]{2019MNRAS.486.5558S}
{Shultz} M.,  {Rivinius} T.,  {Das} B.,  {Wade} G.~A.,   {Chandra} P.,  2019b,
  \mn@doi [\mnras] {10.1093/mnras/stz1129}, \href
  {https://ui.adsabs.harvard.edu/abs/2019MNRAS.486.5558S} {486, 5558}

\bibitem[\protect\citeauthoryear{{Shultz} et~al.,}{{Shultz}
  et~al.}{2019c}]{2019MNRAS.490..274S}
{Shultz} M.~E.,  et~al., 2019c, \mn@doi [\mnras] {10.1093/mnras/stz2551}, \href
  {https://ui.adsabs.harvard.edu/abs/2019MNRAS.490..274S} {490, 274}

\bibitem[\protect\citeauthoryear{{Shulyak}, {Tsymbal}, {Ryabchikova},
  {St{\"u}tz}  \& {Weiss}}{{Shulyak} et~al.}{2004}]{2004A&A...428..993S}
{Shulyak} D.,  {Tsymbal} V.,  {Ryabchikova} T.,  {St{\"u}tz} C.,   {Weiss}
  W.~W.,  2004, \mn@doi [\aap] {10.1051/0004-6361:20034169}, \href
  {https://ui.adsabs.harvard.edu/abs/2004A&A...428..993S} {428, 993}

\bibitem[\protect\citeauthoryear{{Slettebak}}{{Slettebak}}{1955}]{1955ApJ...121..653S}
{Slettebak} A.,  1955, \mn@doi [\apj] {10.1086/146031}, \href
  {https://ui.adsabs.harvard.edu/abs/1955ApJ...121..653S} {121, 653}

\bibitem[\protect\citeauthoryear{{Song} et~al.,}{{Song}
  et~al.}{2022}]{2022A&A...657A..60S}
{Song} H.~F.,  et~al., 2022, \mn@doi [\aap] {10.1051/0004-6361/202141512},
  \href {https://ui.adsabs.harvard.edu/abs/2022A&A...657A..60S} {657, A60}

\bibitem[\protect\citeauthoryear{{Subramanian}, {Balsara}, {ud-Doula}  \&
  {Gagn{\'e}}}{{Subramanian} et~al.}{2022}]{2022MNRAS.515..237S}
{Subramanian} S.,  {Balsara} D.~S.,  {ud-Doula} A.,   {Gagn{\'e}} M.,  2022,
  \mn@doi [\mnras] {10.1093/mnras/stac1778}, \href
  {https://ui.adsabs.harvard.edu/abs/2022MNRAS.515..237S} {515, 237}

\bibitem[\protect\citeauthoryear{{Tetzlaff}, {Neuh{\"a}user}  \&
  {Hohle}}{{Tetzlaff} et~al.}{2011}]{2011MNRAS.410..190T}
{Tetzlaff} N.,  {Neuh{\"a}user} R.,   {Hohle} M.~M.,  2011, \mn@doi [\mnras]
  {10.1111/j.1365-2966.2010.17434.x}, \href
  {https://ui.adsabs.harvard.edu/abs/2011MNRAS.410..190T} {410, 190}

\bibitem[\protect\citeauthoryear{{Townsend}, {Oksala}, {Cohen}, {Owocki}  \&
  {ud-Doula}}{{Townsend} et~al.}{2010}]{2010ApJ...714L.318T}
{Townsend} R.~H.~D.,  {Oksala} M.~E.,  {Cohen} D.~H.,  {Owocki} S.~P.,
  {ud-Doula} A.,  2010, \mn@doi [\apjl] {10.1088/2041-8205/714/2/L318}, \href
  {https://ui.adsabs.harvard.edu/abs/2010ApJ...714L.318T} {714, L318}

\bibitem[\protect\citeauthoryear{{Townsend} et~al.,}{{Townsend}
  et~al.}{2013}]{2013ApJ...769...33T}
{Townsend} R.~H.~D.,  et~al., 2013, \mn@doi [\apj]
  {10.1088/0004-637X/769/1/33}, \href
  {https://ui.adsabs.harvard.edu/abs/2013ApJ...769...33T} {769, 33}

\bibitem[\protect\citeauthoryear{{Vidotto}, {Jardine}, {Morin}, {Donati},
  {Opher}  \& {Gombosi}}{{Vidotto} et~al.}{2014}]{2014MNRAS.438.1162V}
{Vidotto} A.~A.,  {Jardine} M.,  {Morin} J.,  {Donati} J.~F.,  {Opher} M.,
  {Gombosi} T.~I.,  2014, \mn@doi [\mnras] {10.1093/mnras/stt2265}, \href
  {https://ui.adsabs.harvard.edu/abs/2014MNRAS.438.1162V} {438, 1162}

\bibitem[\protect\citeauthoryear{{Vink}, {de Koter}  \& {Lamers}}{{Vink}
  et~al.}{1999}]{1999A&A...350..181V}
{Vink} J.~S.,  {de Koter} A.,   {Lamers} H.~J.~G.~L.~M.,  1999, \aap, \href
  {https://ui.adsabs.harvard.edu/abs/1999A&A...350..181V} {350, 181}

\bibitem[\protect\citeauthoryear{{Vink}, {de Koter}  \& {Lamers}}{{Vink}
  et~al.}{2000}]{2000A&A...362..295V}
{Vink} J.~S.,  {de Koter} A.,   {Lamers} H.~J.~G.~L.~M.,  2000, \aap, \href
  {https://ui.adsabs.harvard.edu/abs/2000A&A...362..295V} {362, 295}

\bibitem[\protect\citeauthoryear{{Vink}, {de Koter}  \& {Lamers}}{{Vink}
  et~al.}{2001}]{2001A&A...369..574V}
{Vink} J.~S.,  {de Koter} A.,   {Lamers} H.~J.~G.~L.~M.,  2001, \mn@doi [\aap]
  {10.1051/0004-6361:20010127}, \href
  {https://ui.adsabs.harvard.edu/abs/2001A&A...369..574V} {369, 574}

\bibitem[\protect\citeauthoryear{{Wade} et~al.,}{{Wade}
  et~al.}{2006}]{2006A&A...458..569W}
{Wade} G.~A.,  et~al., 2006, \mn@doi [\aap] {10.1051/0004-6361:20054759}, \href
  {https://ui.adsabs.harvard.edu/abs/2006A&A...458..569W} {458, 569}

\bibitem[\protect\citeauthoryear{{Washimi} \& {Shibata}}{{Washimi} \&
  {Shibata}}{1993}]{1993MNRAS.262..936W}
{Washimi} H.,  {Shibata} S.,  1993, \mn@doi [\mnras] {10.1093/mnras/262.4.936},
  \href {https://ui.adsabs.harvard.edu/abs/1993MNRAS.262..936W} {262, 936}

\bibitem[\protect\citeauthoryear{{Weber} \& {Davis}}{{Weber} \&
  {Davis}}{1967}]{1967ApJ...148..217W}
{Weber} E.~J.,  {Davis} Leverett J.,  1967, \mn@doi [\apj] {10.1086/149138},
  \href {https://ui.adsabs.harvard.edu/abs/1967ApJ...148..217W} {148, 217}

\bibitem[\protect\citeauthoryear{{ud-Doula} \& {Owocki}}{{ud-Doula} \&
  {Owocki}}{2002}]{2002ApJ...576..413U}
{ud-Doula} A.,  {Owocki} S.~P.,  2002, \mn@doi [\apj] {10.1086/341543}, \href
  {https://ui.adsabs.harvard.edu/abs/2002ApJ...576..413U} {576, 413}

\bibitem[\protect\citeauthoryear{{ud-Doula}, {Owocki}  \&
  {Townsend}}{{ud-Doula} et~al.}{2008}]{2008MNRAS.385...97U}
{ud-Doula} A.,  {Owocki} S.~P.,   {Townsend} R. H.~D.,  2008, \mn@doi [\mnras]
  {10.1111/j.1365-2966.2008.12840.x}, \href
  {https://ui.adsabs.harvard.edu/abs/2008MNRAS.385...97U} {385, 97}

\bibitem[\protect\citeauthoryear{{ud-Doula}, {Owocki}  \&
  {Townsend}}{{ud-Doula} et~al.}{2009}]{2009MNRAS.392.1022U}
{ud-Doula} A.,  {Owocki} S.~P.,   {Townsend} R. H.~D.,  2009, \mn@doi [\mnras]
  {10.1111/j.1365-2966.2008.14134.x}, \href
  {https://ui.adsabs.harvard.edu/abs/2009MNRAS.392.1022U} {392, 1022}

\bibitem[\protect\citeauthoryear{{van Leeuwen}}{{van
  Leeuwen}}{2007}]{2007A&A...474..653V}
{van Leeuwen} F.,  2007, \mn@doi [\aap] {10.1051/0004-6361:20078357}, \href
  {https://ui.adsabs.harvard.edu/abs/2007A&A...474..653V} {474, 653}

\makeatother
\end{thebibliography}




\appendix

\section{The magnetic field components}

Figures A1 through A4 show the strength of the magnetic field components for $\sigma$ Ori E, CU Vir, $\tau$ Sco, and 36 Lyn, respectively.

\begin{figure}
	\includegraphics[width=\columnwidth]{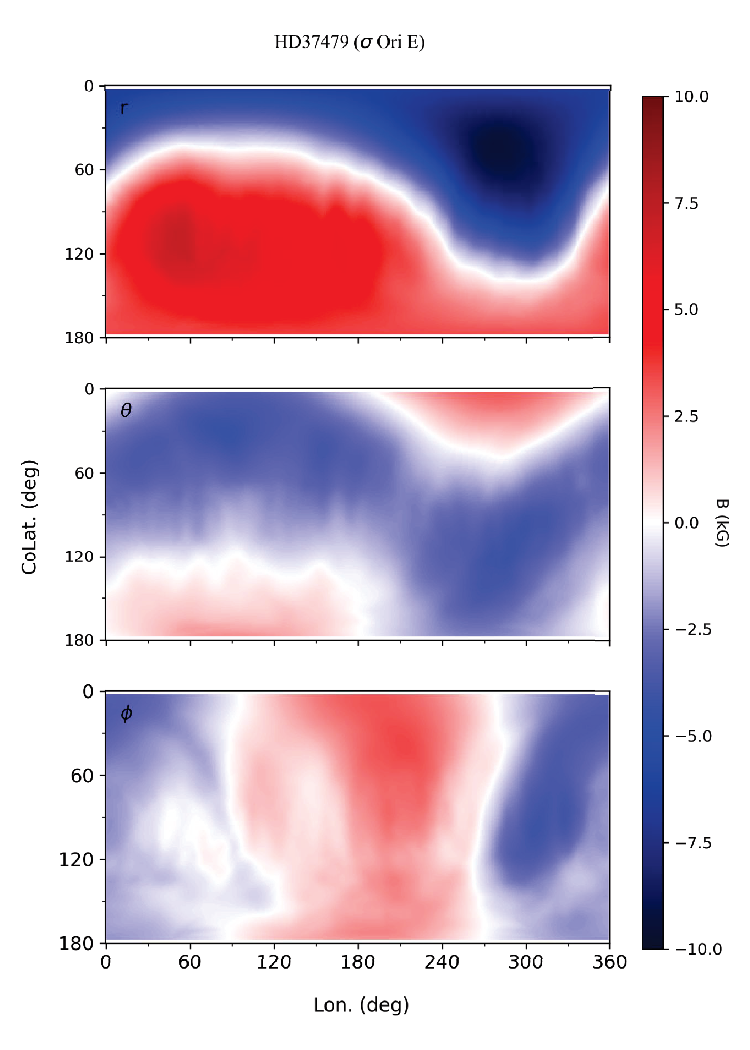}
    \caption{color plots of the strengths of the magnetic field components for $\sigma$ Ori E.}
    \label{fig:figureA1}
\end{figure}

\begin{figure}
	\includegraphics[width=\columnwidth]{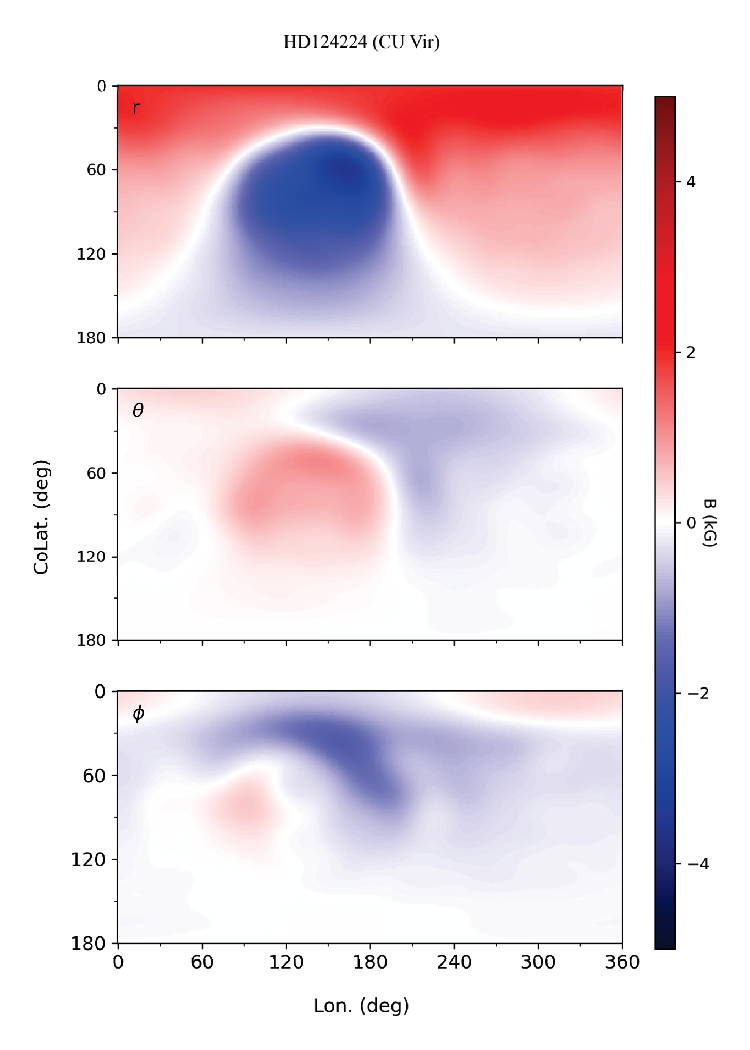}
    \caption{color plots of the strengths of the magnetic field components for CU Vir.}
    \label{fig:figureA2}
\end{figure}

\begin{figure}
	\includegraphics[width=\columnwidth]{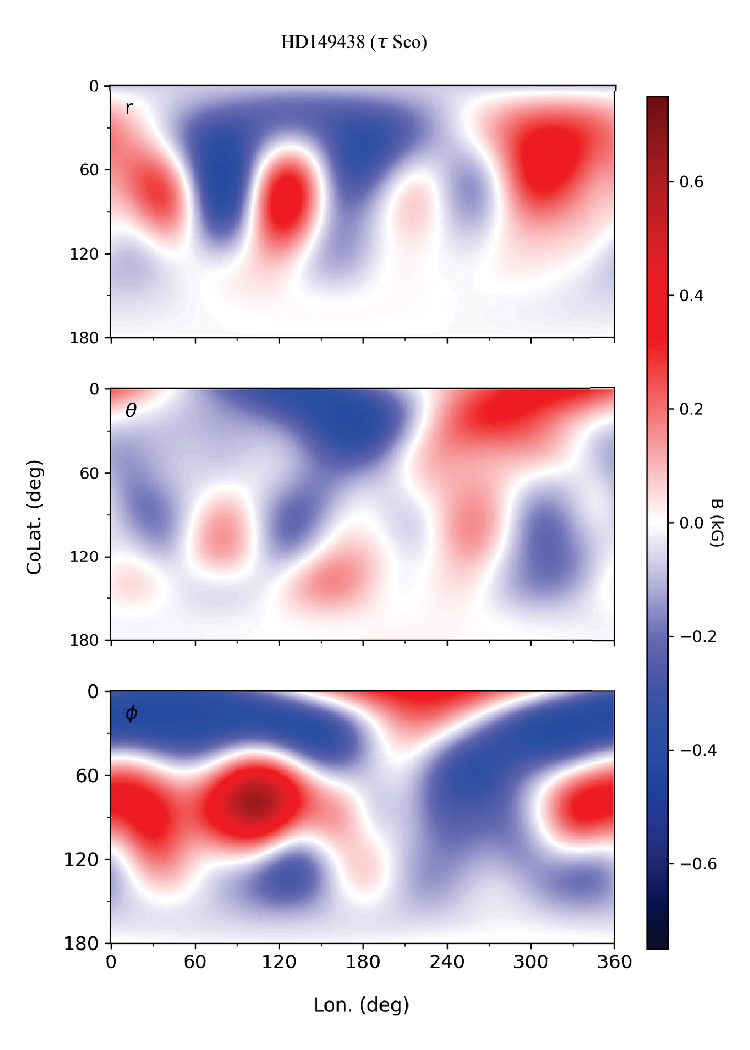}
    \caption{color plots of the strengths of the magnetic field components for $\tau$ Sco.}
    \label{fig:figureA3}
\end{figure}

\begin{figure}
	\includegraphics[width=\columnwidth]{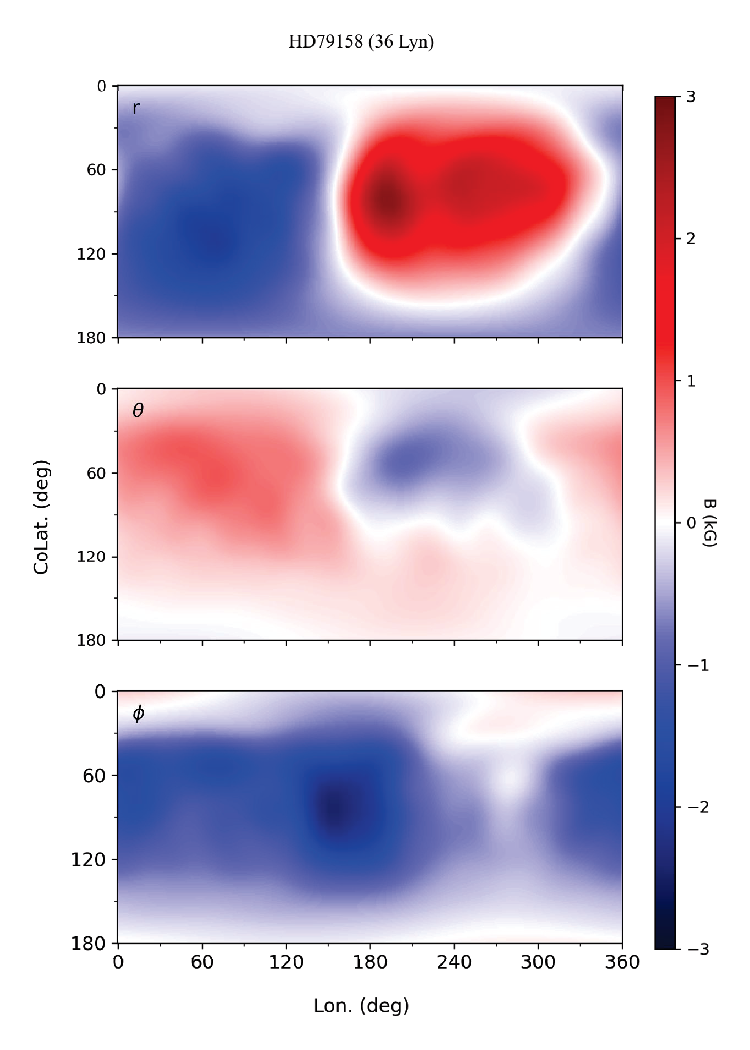}
    \caption{color plots of the strengths of the magnetic field components for 36 Lyn.}
    \label{fig:figureA4}
\end{figure}

Figures A5 through A8 show heat-maps of the $\alpha$, $\beta$, and $\gamma$ harmonic coefficients in equations (1) -(3) from \cite{2014A&A...565A..83K} and the cumulative torque (equation (\ref{eq:13})) histogram of the surface magnetic field for $\sigma$ Ori E, CU Vir, $\tau$ Sco, and 36 Lyn, respectively. For each figure, corresponding to each star, the bottom panel shows a cumulative histogram of the torque computed by reconstructing the ZDI map using only the $\ell$-th harmonics. The horizontal dashed line shows the torque computed from the map reconstructed from all of the harmonics, which is of course the convergence point of the cumulative histogram. To make the figures more compact, the histogram bars are colored according to their sign, with positive/negative in red/blue. 

In a case where the total torque would be reached within a few harmonics only, it could either mean that (i) the small-scale structures are not contributing significantly to the torque or (ii) the field does not have significant small-scale structures to start with. 

\begin{figure}
	\includegraphics[width=\columnwidth]{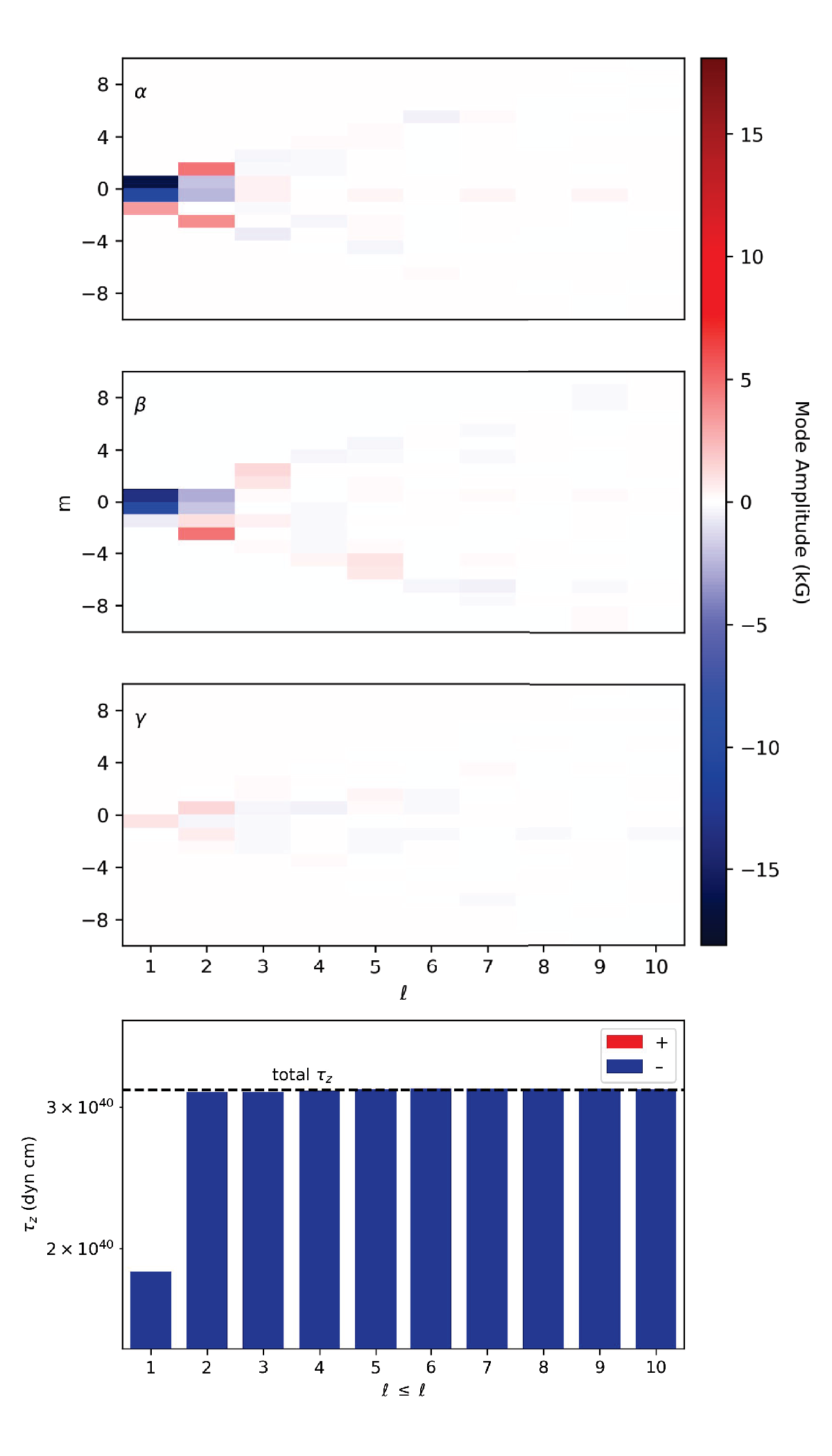}
    \caption{Heat-map of the $\alpha$, $\beta$, and $\gamma$ harmonic coefficients (top) and the cumulative torque histogram of $\sigma$ Ori E.}
    \label{fig:figureA5}
\end{figure}

For $\sigma$ Ori E (fig. A5), high mode amplitude values are clustered around low $\ell$ coefficients indicating that these terms have the highest impact on the overall topography of the surface magnetic field. This is echoed in the cumulative torque histogram which converges to the total torque from the field very quickly ($\ell \leq 2$).

\begin{figure}
	\includegraphics[width=\columnwidth]{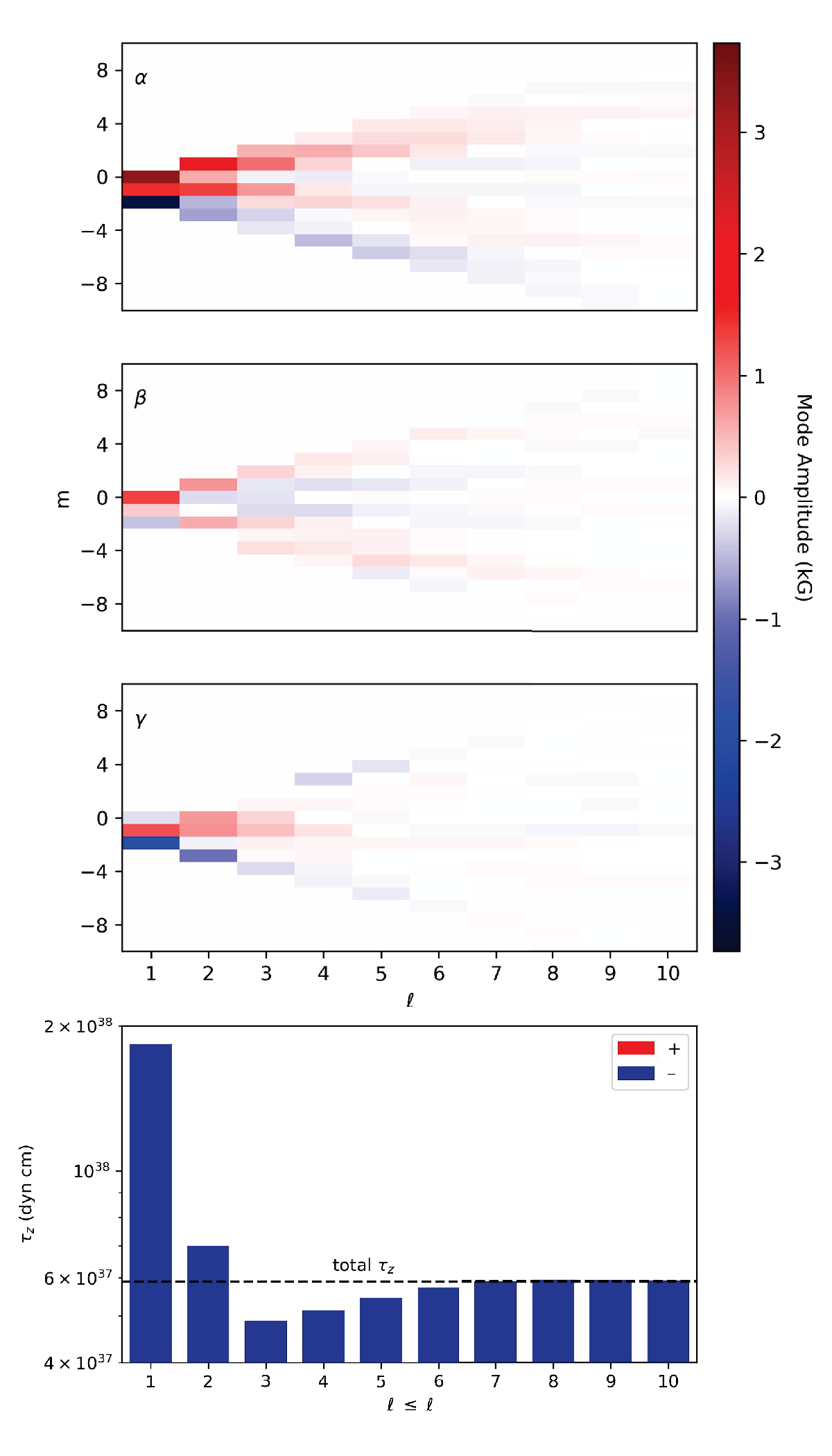}
    \caption{Heat-map of the $\alpha$, $\beta$, and $\gamma$ harmonic coefficients (top) and the cumulative torque histogram of CU Vir.}
    \label{fig:figureA6}
\end{figure}

For CU Vir (fig. A6), high mode amplitude values are found from low $\ell$ to mid - $\ell$ values (out of the total 10) indicating that the overall topography of the surface magnetic field is dominated equally by large and medium-scale structures. This can be seen in the cumulative torque histogram below which converges around $\ell \leq 4$.

\begin{figure}
	\includegraphics[width=\columnwidth]{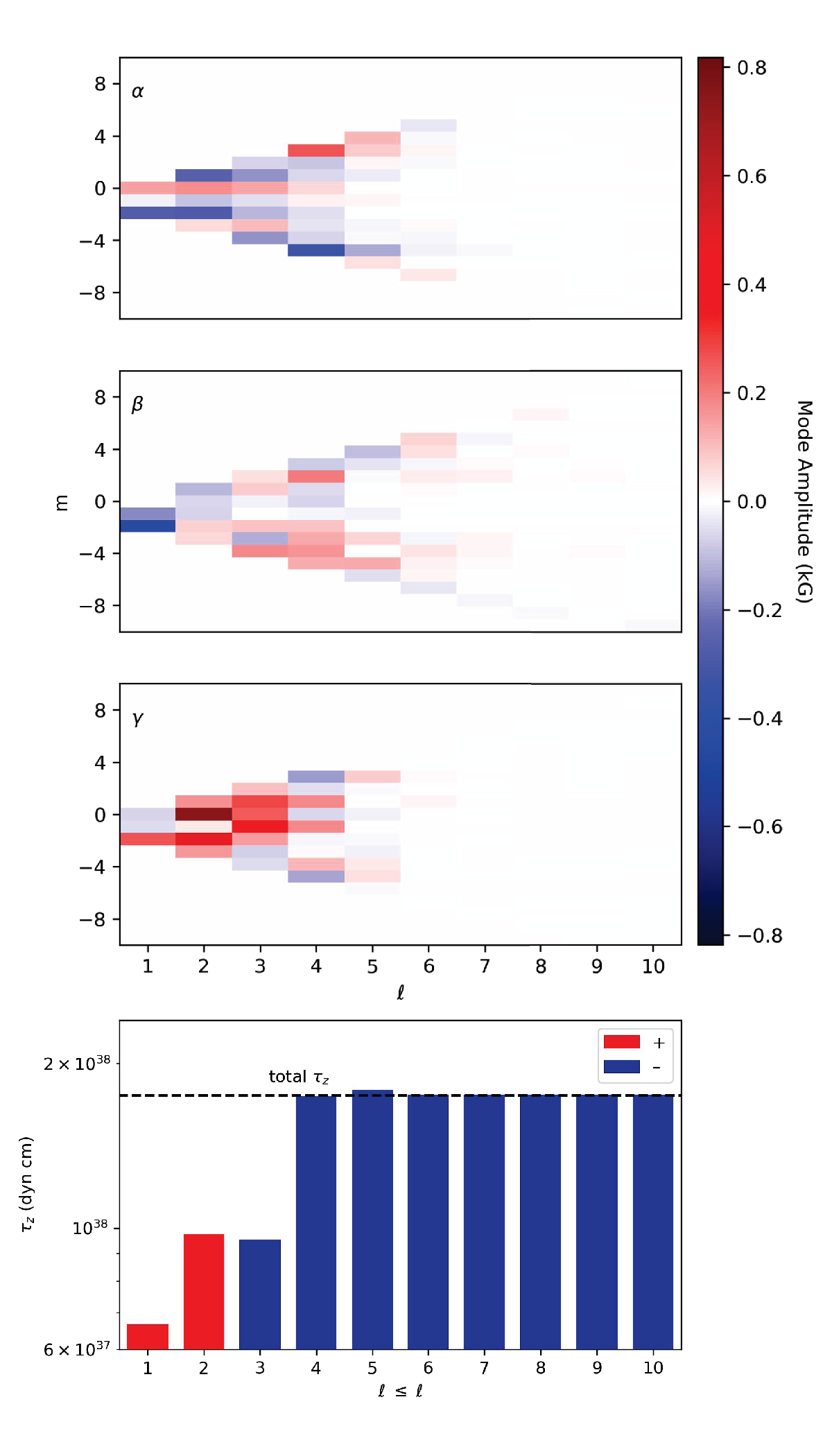}
    \caption{Heat-map of the $\alpha$, $\beta$, and $\gamma$ harmonic coefficients (top) and the cumulative torque histogram of $\tau$ Sco.}
    \label{fig:figureA7}
\end{figure}

For $\tau$ Sco (fig. A7), high mode amplitude values are mostly constrained to low $\ell$ values but extend to higher $\ell$ values for the $\alpha$ terms. This shows that the overall topography of the surface magnetic field is dominated by large and medium-scale structures. This can be seen in the cumulative torque histogram below which converges around $\ell \leq 7$.

\begin{figure}
	\includegraphics[width=\columnwidth]{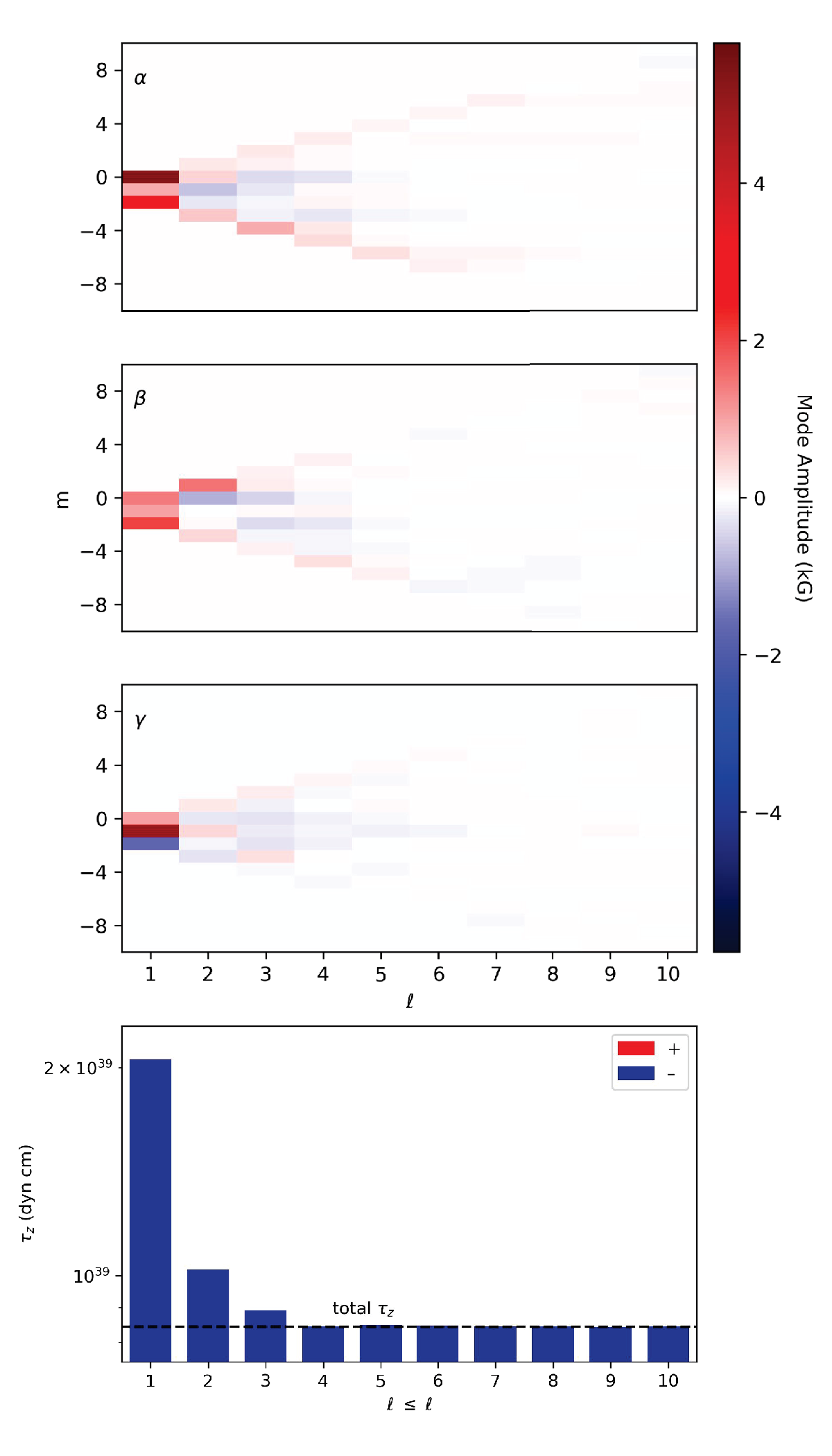}
    \caption{Heat-map of the $\alpha$, $\beta$, and $\gamma$ harmonic coefficients (top) and the cumulative torque histogram of 36 Lyn.}
    \label{fig:figureA8}
\end{figure}

For 36 Lyn (fig. A8), high mode amplitude values are found mostly around low values indicating that the overall topography of the surface magnetic field can be described primarily by large structures. This can be seen in the cumulative torque histogram below which converges around $\ell \leq 4$.


\bsp	
\label{lastpage}
\end{document}